\documentclass[aps,twocolumn,superscriptaddress,floatfix]{revtex4-1}

\usepackage{booktabs}
\usepackage{lipsum}
\usepackage{graphicx}
\usepackage{amsmath,amssymb}
\usepackage{braket}
\usepackage{mathtools}
\usepackage[colorlinks,linkcolor=blue,citecolor=blue,urlcolor=blue]{hyperref}
\usepackage{color}
\usepackage[normalem]{ulem}
\usepackage{appendix,stmaryrd}
\usepackage[thicklines]{cancel}
\usepackage{xcolor}

\newcommand{\be}{\begin{equation}}
\newcommand{\ee}{\end{equation}}

\usepackage{tikz}
\usetikzlibrary{shapes,calc}
\usetikzlibrary{shapes.geometric,arrows.meta,decorations.markings}

\tikzset{
	dot/.style={draw,circle,inner sep=1.5pt,fill=black},
	empty dot/.style={draw,circle,inner sep=1.5pt,fill=white},
	mid arrow/.style={postaction={decorate,decoration={
        markings,
        mark=at position .6 with {\arrow[#1,scale=1.5]{latex}}
    }}},
	spinA/.style={draw=black,thick,circle,inner sep=2.5pt, fill=figBlue},
	spinBC/.style={draw=black,thick,circle,inner sep=2.5pt, fill=figRed},
	faded/.style={opacity=0.2},
}

\begin{document}

\title{Electrical magnetochiral anisotropy and quantum metric in chiral conductors}

\author{Yiyang Jiang}
\affiliation{Department of Condensed Matter Physics, Weizmann Institute of Science, Rehovot 7610001, Israel}
\author{Qinyan Yi}
\affiliation{Department of Condensed Matter Physics, Weizmann Institute of Science, Rehovot 7610001, Israel}
\author{Binghai Yan}
\email[]{binghai.yan@weizmann.ac.il}
\affiliation{Department of Condensed Matter Physics, Weizmann Institute of Science, Rehovot 7610001, Israel}
\date{\today}

\begin{abstract}
Electrical magnetochiral anisotropy (EMCA) refers to the chirality- and current-dependent nonlinear magnetoresistance in chiral conductors and is commonly interpreted in a semiclassical picture. 
In this work, we reveal a quantum geometry origin of EMCA using a chiral rectangular lattice model that resembles a chiral organic conductor (DM-EDT-TTF)${}_2$ClO${}_4$ studied for EMCA recently and exhibits symmetry-protected Dirac bands similar to those of graphene. 
Compared to the semiclassical term, we find that Dirac states contribute significantly to both traditional longitudinal EMCA and the unconventional transverse EMCA via the quantum metric when Fermi energy is close to the Dirac point. 
Besides, we discover that a topological insulator state can emerge once spin-orbit coupling (SOC) is added to our chiral model lattice. 
Our work paves a path toward understanding quantum geometry in the magnetotransport of chiral materials. 
\end{abstract}
\maketitle


\section{Introduction}\label{intro}

Chirality leads to exotic nonreciprocal phenomena in transport and optics in chiral materials \cite{tokura2018nonreciprocal, rikken1997observation, krichevtsov1993spontaneous, nagaosa2017naturephysics, Naaman2022, Yan2024}. Among them, electrical magnetochiral anisotropy (EMCA)~\cite{rikken2001} refers to a magnetoresistance that depends linearly on the current ($\mathbf{I}$), magnetic field ($\mathbf{B}$), and the chirality ($\chi=\pm$).
EMCA was experimentally observed in many chiral crystalline materials such as chiral Bi wires~\cite{rikken2001}, chiral carbon nanotubes \cite{krstic2002magneto}, organic conductors \cite{pop2014EMCA}, and chiral crystals (e.g. Te) \cite{rikken2019Te, calavalle2022gate, inui2020chirality, shiota2021chirality, niu2023tunable}. Most of these materials are crystalline and thus can be described by band theory. 

\begin{figure*}[tbp]
\includegraphics[width=0.95\linewidth]{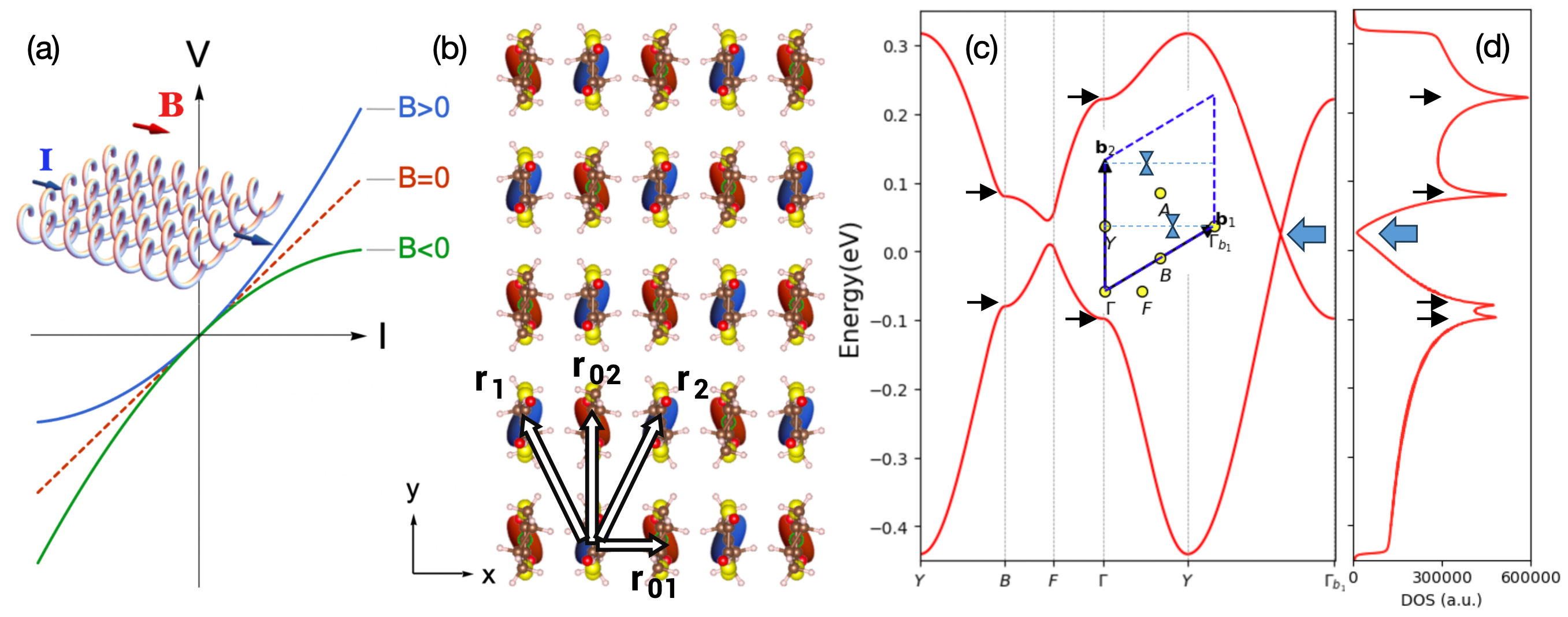}
\caption{ (a) Typical I-V curve of the EMCA effect. The chiral lattice exhibits chirality ($\chi=\pm$, representing left- or right-handed enantiomers) and current-dependent magnetoresistance, which can be phenomenologically described by $R^{\chi}(\mathbf{I}; \mathbf{B}) = R_0 \left(1 + \beta B^2 + \gamma^{\chi} \mathbf{B} \cdot \mathbf{I}\right)$, where $R_0$ represents the normal linear resistivity, $\beta$ is the normal magnetoresistance coefficient, and $\gamma^\chi$ stands for the EMCA coefficient satisfying $\gamma^{-\chi} = - \gamma^\chi$. The inset shows a model setup for EMCA measurement on a chiral crystal where $\mathbf{B} \parallel \mathbf{I}$.
(b) Top view of the chiral lattice model. In the donor layer of (DM-EDT-TTF)$_2$ClO$_4$, the red and blue ellipsoids represent two types of chiral enantiomers that individually form triangular lattices. The two sublattices are separated in $z$-direction, breaking inversion and mirror symmetries.
(c,d) The band structure and density of states (DOS) of the chiral lattice model with parameters $t_0 = 0.135\ \text{eV}$, $t_1 = -0.055\ \text{eV}$, $t_2 = 0.015\ \text{eV}$, $\Delta t_2 = 0.02\ \text{eV}$, $\delta U = \mu = 0\ \text{eV}$. The inset in (c) shows the 2D Brillouin zone. The blue and black arrows indicate Dirac points and van Hove singularities in the band structure.}
\label{fig1}
\end{figure*}

A typical EMCA setup with $\mathbf{B||I}$ and resultant nonlinear current-voltage (I-V) curves is drawn in Fig. \ref{fig1} (a). 
Phenomenologically, $\mathbf{I}$ generates a $\chi$-dependent self-magnetic field in a chiral conductor and induces a negative or positive magnetoresistance, depending on the parallel or antiparallel alignment between the self-field and external magnetic field. The total resistance then depends on the current direction, leading to a nonlinear I-V relationship. 
Semiclassically, the chiral structure of materials imparts electrons with chiral orbitals~\cite{liu2021chirality} that shows $\mathbf{I}$- and $\chi$-dependent net orbital angular momentum (OAM). When this OAM couples to $\mathbf{B}$~\cite{rikken2019Te, nagaosa2017naturephysics, Liu2021UMR}, the band dispersion tilts, giving rise to EMCA. Both mechanisms originate from the coupling between chiral motions of electrons and the external magnetic field, which leads to the $\gamma^{\chi} \mathbf{B} \cdot \mathbf{I}$ term in the resistance. Nevertheless, only the band dispersion is considered here, while the wavefunction information that encodes the quantum geometry is neglected.

On the other hand, the wavefunction-associated quantum geometry has received great attention in modern condensed matter physics~\cite{provost1980riemannian, anandan1990geometry, xiao2010berry} and can lead to exotic geometric phenomena, including nonlinear transport~\cite{Gao2014field, sodemann2015quantum, wang2023quantum, gao2023quantum, daniel2023unifying, daniel2024unification, gao2019semiclassical, ma2019observation}. In addition to the dispersion-related Drude term, a novel intrinsic quantum metric contribution has been predicted~\cite{Gao2014field, daniel2024unification} and observed experimentally~\cite{wang2023quantum, gao2023quantum}. This nonlinear response theory was developed for intrinsic magnetic solids because the quantum metric contribution requires time-reversal symmetry ($\mathcal{T}$) breaking. While almost all EMCA samples are non-magnetic~\cite{rikken2001, krstic2002magneto, pop2014EMCA, rikken2019Te, calavalle2022gate, inui2020chirality, shiota2021chirality, niu2023tunable}, the external magnetic field inevitably breaks $\mathcal{T}$ symmetry, thus creating a potential playground to examine the quantum metric effect.

In this article, we demonstrate the quantum geometric contribution to EMCA using a chiral lattice model that resembles the chiral conductor (DM-EDT-TTF)$_2$ClO$_4$, studied for EMCA~\cite{pop2014EMCA}. We incorporate the magnetic field into the lattice model via a gauge field to investigate the quantum metric contribution to the nonlinear conductivity. The chiral lattice exhibits symmetry-protected Dirac points in the band structure, similar to graphene. Through numerical calculations, we observe a significant geometric EMCA generated by Dirac states when the chemical potential is near the Dirac point. Additionally, we demonstrate that a transverse EMCA can generally exist and exhibits substantial enhancement in the vicinity of the Dirac cone, challenging the traditional view of EMCA as a purely longitudinal effect. Furthermore, we discovered that a topological insulator state can emerge when a large spin-orbit coupling (SOC) is added to our chiral lattice model, which resembles the Kane-Mele model~\cite{kane2005qsh, kane2005z2}. Our work reveals a potential platform for demonstrating quantum geometric effects in magnetotransport arising from the topological band structure of nonmagnetic chiral metals, thereby expanding the potential for designing advanced magnetotransport and rectifier devices that harness this intrinsic nonreciprocal effect.

\section{Lattice Symmetry and Tight-binding Model}

\subsection{Lattice Symmetry}

We construct a chiral lattice model that has the same symmetry as the donor layer of the chiral salt (DM-EDT-TTF)$_2$ClO$_4$, studied for EMCA~\cite{pop2014EMCA}, where charge transport occurs mainly within the layer.
The lattice of the chiral material consists of two types of chiral enantiomers, represented by red and blue ellipsoids in the lattice, as shown in Fig.~\ref{fig1}(b). Each type of enantiomer forms a triangular lattice with lattice constant $a$. The two sets of triangular lattices are separated in the $x$ and $z$ directions by $\frac{a}{2}$ and $\Delta z$, respectively, resulting in a chiral 2D lattice. The symmetry of the chiral lattice model belongs to layer group $p112$ (No.~3), which is generated by a two-fold rotation about the $z$ axis ($C_{2z}$) and time-reversal symmetry $\mathcal{T}$. As we will see, the combined symmetry $C_{2z}\mathcal{T}$ is crucial for protecting Dirac states, similar to the case of graphene.

\subsection{Tight-binding Model}\label{sec1}

We first construct a spinless tight-binding model with a minimal set of hopping parameters: \( t_0 \) for nearest-neighbor (N.N.) hopping along \( \mathbf{r}_{01} = \left(\frac{1}{2}, 0\right)a \), \( t_1 \) for next-nearest-neighbor (N.N.N.) hopping along \( \mathbf{r}_{02} = \left(0, \frac{\sqrt{3}}{2}\right)a \), and, most importantly, a set of anisotropic next-next-nearest-neighbor (N.N.N.N.) hoppings \( t_2 \pm \Delta t_2 \) along \( \mathbf{r}_{1,2} = \mp \mathbf{r}_{01} + \mathbf{r}_{02} \), which induces chirality in the model. Specifically, the red triangular lattice favors hopping in the \( \mathbf{r}_1 \) direction (with hopping amplitude \( t_2 + \Delta t_2 \)) but disfavors hopping in the \( \mathbf{r}_2 \) direction (with hopping amplitude \( t_2 - \Delta t_2 \)), while the blue lattice has exactly the opposite hopping preferences. We can also introduce additional parameters like the Fermi level \( \mu \) and the sublattice potential \( \delta U \) to enrich our model. Because of the negligible spin-orbit coupling (SOC) in this organic material, we neglect the spin in the discussion of EMCA and will reconsider it for possible topological states later.

The spinless Bloch Hamiltonian has the form:
\begin{equation}
H(\mathbf{k}) = d_0(\mathbf{k}) \sigma_0 + \mathbf{d}(\mathbf{k}) \cdot \boldsymbol{\sigma},
\end{equation}
where the Pauli matrices \( \boldsymbol{\sigma} \) represent the sublattice degree of freedom in the Bloch Hamiltonian. Using the hopping parameters mentioned above, it's straightforward to obtain the coefficients for \( \sigma_i \):
\begin{equation}
\begin{aligned}
& d_0(\mathbf{k}) = 2 t_2 \left[\cos(\mathbf{k} \cdot \mathbf{r}_2) + \cos(\mathbf{k} \cdot \mathbf{r}_1)\right] - \mu, \\ 
& d_x(\mathbf{k}) = 2 t_0 \cos(\mathbf{k} \cdot \mathbf{r}_{01}) + 2 t_1 \cos(\mathbf{k} \cdot \mathbf{r}_{02}), \\
& d_y(\mathbf{k}) = 0, \\
& d_z(\mathbf{k}) = 2 \Delta t_2 \left[\cos(\mathbf{k} \cdot \mathbf{r}_2) - \cos(\mathbf{k} \cdot \mathbf{r}_1)\right] + \delta U.
\end{aligned}
\label{Hamiltonian}
\end{equation}
Here, we emphasize that the vanishing \( d_y(\mathbf{k}) \) is due to the \( C_{2z} \) and \( \mathcal{T} \) symmetries. By symmetry analysis, the symmetry operators in the spinless model are \( C_{2z} = - \sigma_0 \) and \( \mathcal{T} = \sigma_0 \mathcal{K} \), where \( \mathcal{K} \) denotes complex conjugation. The \( C_{2z} \) and \( \mathcal{T} \) symmetries enforce \( H(\mathbf{k}) = H(-\mathbf{k}) \) and \( H^*(\mathbf{k}) = H(-\mathbf{k}) \), respectively. Since only \( \sigma_y^* = -\sigma_y \), the vanishing of \( d_y(\mathbf{k}) \) is constrained by the \( C_{2z} \) and \( \mathcal{T} \) symmetries.

The dispersion of the two-band system is \( E_{\pm}(\mathbf{k}) = d_0(\mathbf{k}) \pm |\mathbf{d}(\mathbf{k})| \). In a 2D system with two momentum components \( k_x \) and \( k_y \), finding a gapless point typically requires satisfying three equations \( d_i(k_x, k_y) = 0 \) for \( i = x, y, z \). Therefore, a Dirac cone is generally accidental in 2D lattices. However, if symmetry enforces \( d_y(\mathbf{k}) = 0 \), then it is possible to find robust Dirac cones for 2D systems with a two-dimensional band representation, similar to the case of graphene.

Indeed, we found \( C_{2z}\mathcal{T} \)-protected Dirac cones in both the tight-binding model and density-functional theory (DFT) calculations (Figs.~\ref{fig1} and \ref{appfig1}). However, unlike graphene, whose Dirac points are pinned at the \( K/K' \) points by an extra \( C_{3z} \) rotational symmetry, our Dirac cones are not pinned at any high-symmetry points and are generally located along the \( \Gamma \)-\( Y \) line, as shown in the inset of Fig.~\ref{fig1}(c). Two Dirac points exist here due to the fermion doubling theorem~\cite{nielsen1981absence}. We note that the real material's chemical potential is located about 0.18~eV above the Dirac point in DFT calculations. More details of the real material's band dispersion are shown in Fig.~\ref{appfig1} in Appendix~\ref{app0}.

Specifically, we can include the magnetic field \( \mathbf{B} \) into our system by the Peierls substitution~\cite{blount1962bloch}. Since \( \mathbf{B} \) is in-plane, we can consider a vector potential that maintains 2D translational symmetry, \( \mathbf{A}(\mathbf{r}) = \mathbf{B} \times \mathbf{r} \). Then the spinless Bloch Hamiltonian is transformed into
\begin{equation}
H(\mathbf{k}; \mathbf{B}) = d_0(\mathbf{k}; \mathbf{B}) \sigma_0 + \mathbf{d}(\mathbf{k}; \mathbf{B}) \cdot \boldsymbol{\sigma},
\end{equation}
where
\begin{widetext}
\begin{equation}
\begin{aligned}
d_0(\mathbf{k}; \mathbf{B}) &= 2 t_2 \left[ \cos(\mathbf{k}_B \cdot \mathbf{r}_2) \cos(\mathbf{k} \cdot \mathbf{r}_2) + \cos(\mathbf{k}_B \cdot \mathbf{r}_1) \cos(\mathbf{k} \cdot \mathbf{r}_1) \right] \\
&\quad + 2 \Delta t_2 \left[ \sin(\mathbf{k}_B \cdot \mathbf{r}_1) \sin(\mathbf{k} \cdot \mathbf{r}_1) - \sin(\mathbf{k}_B \cdot \mathbf{r}_2) \sin(\mathbf{k} \cdot \mathbf{r}_2) \right] - \mu, \\
d_x(\mathbf{k}; \mathbf{B}) &= 2 t_0 \cos(\mathbf{k} \cdot \mathbf{r}_{01}) + 2 t_1 \cos(\mathbf{k} \cdot \mathbf{r}_{02}), \\
d_y(\mathbf{k}; \mathbf{B}) &= 0, \\
d_z(\mathbf{k}; \mathbf{B}) &= 2 \Delta t_2 \left[ \cos(\mathbf{k}_B \cdot \mathbf{r}_2) \cos(\mathbf{k} \cdot \mathbf{r}_2) - \cos(\mathbf{k}_B \cdot \mathbf{r}_1) \cos(\mathbf{k} \cdot \mathbf{r}_1) \right] \\
&\quad - 2 t_2 \left[ \sin(\mathbf{k}_B \cdot \mathbf{r}_1) \sin(\mathbf{k} \cdot \mathbf{r}_1) + \sin(\mathbf{k}_B \cdot \mathbf{r}_2) \sin(\mathbf{k} \cdot \mathbf{r}_2) \right] + \delta U,
\end{aligned}
\label{Hamiltonian_adding_B}
\end{equation}
\end{widetext}
where \( \mathbf{k}_B \equiv \frac{e \Delta z}{\hbar} \mathbf{e}_z \times \mathbf{B} \), and \( \Delta z \) is the separation between the two chiral sublattices in the \( z \) direction. More details about the derivation of tight-binding model and Peierls substitution can be found in App. \ref{appPeierls}.

\subsection{\( C_{2z}\mathcal{T} \)-protected Dirac cones}

When measuring EMCA, the in-plane magnetic field \( \mathbf{B} \) breaks both \( C_{2z} \) and \( \mathcal{T} \) symmetries. However, since \( \mathbf{B} \) transforms to \( -\mathbf{B} \) under both \( C_{2z} \) and \( \mathcal{T} \), the combined space-time symmetry \( C_{2z}\mathcal{T} \) is preserved, which enforces the reality of the Bloch Hamiltonian \( H^*(\mathbf{k}; \mathbf{B}) = H(\mathbf{k}; \mathbf{B}) \). Due to similar symmetry protection reasons as stated in Sec.~\ref{sec1}, we are assured that the robust Dirac cone is still protected by \( C_{2z}\mathcal{T} \) symmetry even with an in-plane magnetic field, since the \( d_y = 0 \) constraint persists~\cite{zhao2024hybrid, fang2015topological}.

After implementing the in-plane magnetic field \( \mathbf{B} \) along the \( y \) direction, the Dirac cones will be tilted along the \( y \) direction.

The tilted band structure in the \( k_y \) direction represents a preferred momentum along the \( y \) direction when the electron undergoes cyclotron motion around the \( y \) axis (caused by \( \mathbf{B} \)). The rotation and the preferred momentum align together to form a helical movement that corresponds to the schematic in the inset of Fig.~\ref{fig1}(a), giving rise to EMCA.

In fact, the position and energy of the Dirac cones also shift when \( \mathbf{B} \) is nonzero. However, a pure positional shift in momentum doesn't affect the total conductivity of the whole Brillouin zone. Also, the shift in the Dirac point energy is a second-order effect in \( \mathbf{B} \) and thus can be neglected in EMCA, which is a linear effect in \( \mathbf{B} \). Therefore, it is the tilting of the band structure that contributes to EMCA.

\section{Nonlinear Conductivity and Calculation of EMCA}
\subsection{EMCA as nonlinear conductivity}

We can describe EMCA in terms of conductivity, \(\sigma\), and current density, \(\mathbf{j}\), as follows:
\begin{equation}
\sigma_{a;a}^{\chi} (\mathbf{j}; \mathbf{B}) = \sigma_0 \left(1 - \beta B^2 - \hat{\gamma}^{\chi} \mathbf{B} \cdot \mathbf{j} \right)
\end{equation}

This formulation is more convenient when discussing its relation to the nonlinear response. Here, we use \(\hat{\gamma}\) to distinguish it from \(\gamma\) in the original magnetoresistance formula, \(R^{\chi}(\mathbf{I}; \mathbf{B}) = R_0 \left(1 + \beta B^2 + \gamma^{\chi} \mathbf{B} \cdot \mathbf{I} \right)\). Since the SI unit for current density is \(\text{A} \cdot \text{m}^{-2}\), the unit for the corresponding EMCA coefficient \(\hat{\gamma}\) is \(\text{m}^2 \text{T}^{-1} \text{A}^{-1}\). We also assume negligible transverse resistivity, and that \(\beta B^2 \ll 1\) and \(\hat{\gamma}^{\chi} \mathbf{B} \cdot \mathbf{j} \ll 1\).

Following the perturbative analysis in Appendix \ref{app1}, the EMCA coefficient can be obtained by calculating the linear and nonlinear DC conductivity, as shown in \cite{nagaosa2017naturephysics}:
\begin{equation}
\hat{\gamma} = - \frac{\left.\partial \sigma^{(2)}_{aa;a}/\partial |\mathbf{B}| \right|_{B=0}}{\left[\left.\sigma^{(1)}_{a;a}(\mathbf{B})\right|_{B=0}\right]^2} = - \lim_{B \to 0} \frac{\sigma^{(2)}_{aa;a}(\mathbf{B})}{|\mathbf{B}| \left[\sigma^{(1)}_{a;a}(\mathbf{B}) \right]^2}
\end{equation}

Here, \(\sigma^{(1)}_{a;a}\) and \(\sigma^{(2)}_{a;aa}\) represent the linear and second-order longitudinal conductivities, respectively, where the index \(a = x, y\) denotes the spatial components of the conductivity tensors. We also point out that a transverse EMCA can also be defined if we generalize the definition. Basically, the most general EMCA coefficients should possess three spatial components, indicating both longitudinal and transverse response.
\begin{equation}
\hat{\gamma}^{ab;c} = - \lim_{B \to 0} \frac{\sigma^{(2)}_{ab;c}(\mathbf{B})}{|\mathbf{B}| \left[\sigma^{(1)}_{a;a}(\mathbf{B}) \right]^2}
\end{equation}

\subsection{Semiclassical Nonlinear Transport Theory}

Semiclassically, the current density of the system is given by:
\begin{equation}
\mathbf{j} = - e \sum_{n} \int_k f_n \mathbf{v}_n
\end{equation}

Here, \(\int_k \equiv \int \frac{d^D \mathbf{k}}{(2\pi)^D}\) is the integral over $D$-dimensional Brillouin zone, \(-e\) is the charge of the electron, and \(f_n\) is the distribution function of the \(n\)-th band, which satisfies the Boltzmann transport equation \cite{daniel2024unification}:
\begin{equation}
\partial_t f - e \frac{ \mathbf{E}}{\hbar} \cdot \nabla_\mathbf{k} f + \mathbf{v} \cdot \nabla_\mathbf{r} f = - \frac{f - f^{(0)}}{\tau}
\end{equation}
where \(\tau\) is the transport relaxation time that can be affected by the strength of many-body interaction effect or impurity scattering, and \(f^{(0)}\) is the Fermi-Dirac distribution. Assuming the DC limit and spatial homogeneity, the steady-state distribution function can be solved perturbatively as \(f_n = f_n^{(0)} + f_n^{(1)} + f_n^{(2)} + \dots\), where \(f_n^{(l)} \approx \left( \frac{e \tau}{\hbar} \mathbf{E} \cdot \nabla_{\mathbf{k}} \right)^l f_n^{(0)}\). Here we also assume the interaction or disorder strength to be a perturbation to the single-particle band picture, because the original material is a good band metal~\cite{pop2014EMCA}.

\(\mathbf{v}_n\) is the group velocity of the \(n\)-th band, which includes both the band dispersion contribution and the anomalous velocity from the equation of motion \cite{xiao2010berry}:
\begin{equation}
\mathbf{v}_n = \frac{1}{\hbar} \frac{\partial \varepsilon_n}{\partial \mathbf{k}} - \frac{e}{\hbar} \mathbf{E} \times \boldsymbol{\Omega}_n
\end{equation}
where \(\varepsilon_n\) is the dispersion of the \(n\)-th band and \(\boldsymbol{\Omega}_n\) is the Berry curvature of the \(n\)-th band. We can also expand \(\mathbf{v}_n\) in orders of \(\mathbf{E}\), i.e., \(\mathbf{v}_n = \mathbf{v}_n^{(0)} + \mathbf{v}_n^{(1)} + \mathbf{v}_n^{(2)} + \dots\), where \(\mathbf{v}_n^{(l)} = \frac{1}{\hbar} \frac{\partial \varepsilon_n^{(l)}}{\partial \mathbf{k}} - \frac{e}{\hbar} \mathbf{E} \times \boldsymbol{\Omega}_n^{(l-1)}\). The \(l\)-th order correction to the dispersion of the \(n\)-th band, \(\varepsilon_n^{(l)}\), and the \((l-1)\)-th order correction to the Berry curvature of the \(n\)-th band, \(\boldsymbol{\Omega}_n^{(l-1)}\), can be obtained from canonical perturbation theory, discussed in more detail in Appendix \ref{app2}. 

This gives the expression for the \(l\)-th order electron current \(\mathbf{j}^{(l)} = -e \int_k \sum_{p=0}^n f_n^{(p)} \mathbf{v}_n^{(l-p)}\),
from which we can obtain the general expressions for the first-order and second-order conductivities \cite{daniel2024unification}:
\begin{equation}
\sigma^{a;b} = - \frac{e^2 \tau}{\hbar^2} \sum_n \int_k f_n^{(0)} \partial_{k_a} \partial_{k_b} \varepsilon_n
\label{eqconductivity0}
\end{equation}
\begin{equation}
\begin{aligned}
& \sigma^{ab;c} = \sigma^{ab;c}_{\text{disp}} + \sigma^{ab;c}_{\text{geo}} \\
& = - \frac{e^3 \tau^2}{\hbar^3} \sum_n \int_k f_n^{(0)} \partial_{k_a} \partial_{k_b} \partial_{k_c} \varepsilon_n \\
& - \frac{e^3}{\hbar} \sum_n \int_k f_n^{(0)} \left[2 \partial_{k_c} G_n^{ab} - \frac{1}{2} \left( \partial_{k_a} G_n^{bc} + \partial_{k_b} G_n^{ac} \right) \right]
\end{aligned}
\label{eqconductivity}
\end{equation}
where \(\sigma^{ab;c}_{\text{disp}}\) and \(\sigma^{ab;c}_{\text{geo}}\) represent the contributions from pure dispersion \(\varepsilon_n\) and the (band-normalized) quantum metric dipole \(\partial_k G_n\), which is a quantum geometric contribution. The quantum metric of the \(n\)-th band is defined as \(G_n^{ab} \equiv \sum_{m \neq n} \left( A_{nm}^a A_{mn}^b + A_{nm}^b A_{mn}^a \right) / \varepsilon_{nm}\), where \(A_{nm}^a\) is the \(a\)-th spatial component of the non-Abelian Berry connection, and \(\varepsilon_{nm} \equiv \varepsilon_n - \varepsilon_m\) is the energy difference between the \(n\)-th and \(m\)-th bands.

Inspired by this, we can separate the EMCA coefficient into two parts, \(\hat{\gamma} = \hat{\gamma}_{\text{disp}} + \hat{\gamma}_{\text{geo}}\), to account for both dispersion and quantum metric contributions to the EMCA effect.

It is worth noting that we neglect the contribution from the Berry curvature term in Eq. \eqref{eqconductivity0} and Eq. \eqref{eqconductivity} compared to the full expression of nonlinear conductivity in Eq. \eqref{eqS} derived in Sec. \ref{secconductivity}. This is because the \(C_{2z}\mathcal{T}\) symmetry of the system forces the Berry curvature to vanish. Moreover, the Berry curvature only contributes to the transverse conductivity, which is outside the scope of our discussion on EMCA. Therefore, the only contributions to EMCA in our model are from the dispersion-dependent \(\hat{\gamma}_{\text{disp}}\) part and the intrinsic quantum metric-associated \(\hat{\gamma}_{\text{geo}}\) part. However, in more general systems with fewer symmetries, multiple terms—including the nonlinear Drude term, Berry curvature dipole, and quantum metric dipole—can all contribute to EMCA.

\begin{figure*}[htbp]
\includegraphics[width=\linewidth]{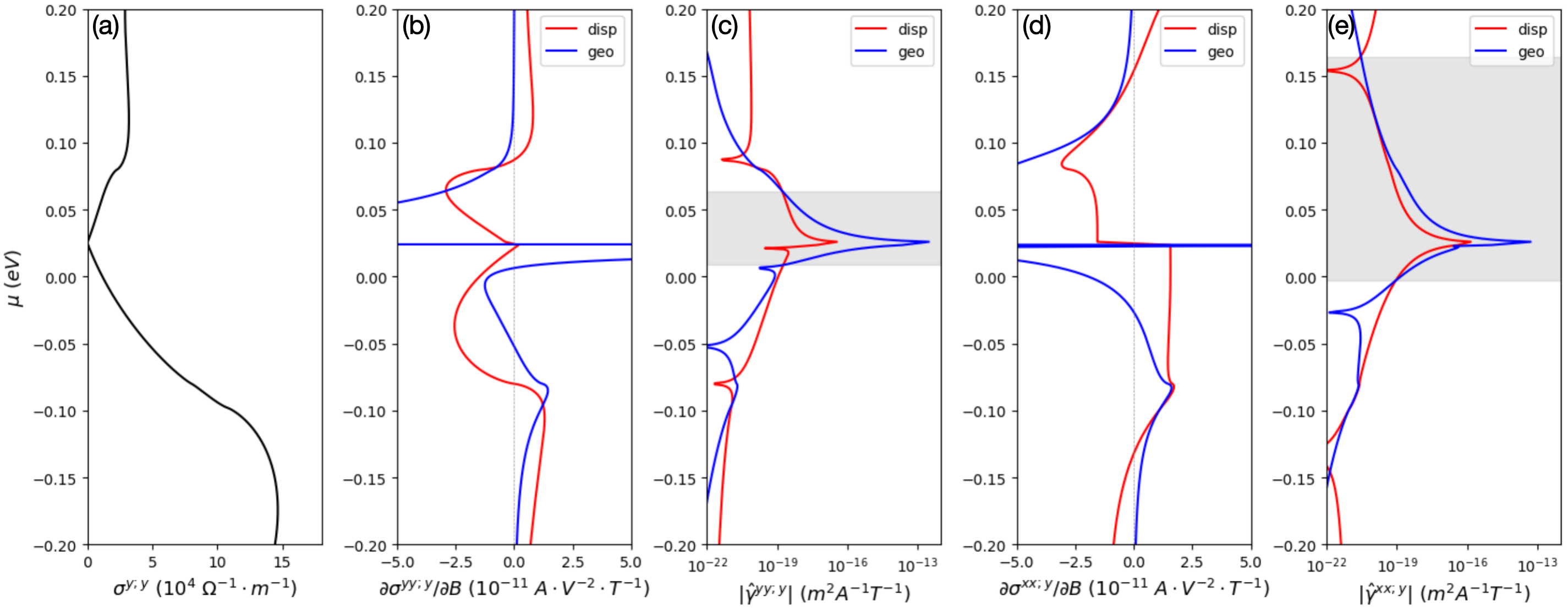}
\caption{ \label{FigS1} (a) First-order longitudinal conductivity along the $y$ direction as a function of Fermi energy. 
(b,d) Dispersion and geometric contributions to the second-order (b) longitudinal and (d) transverse conductivity at different Fermi energies. 
(c,e) Comparison of the order of magnitude between the dispersion contribution, \(\hat{\gamma}_{\text{disp}}\) and the geometric contribution, \(\hat{\gamma}_{\text{geo}}\) of the EMCA coefficient \(\hat{\gamma}\) in (c) longitudinal and (e) transverse EMCA, respectively. The shaded grey region indicates the regime where the EMCA is dominated by the quantum metric contribution. The typical carrier lifetime for organic conductors, \(\tau = 1 \times 10^{-14}\) s, is used~\cite{lifetimenote}.
}\label{fig2}
\end{figure*}

\subsection{Geometric EMCA from the Dirac Cone}

In this section, we will use a simple scaling rule to analyze the different scaling behaviors of \(\hat{\gamma}_{\text{disp}}\) and \(\hat{\gamma}_{\text{geo}}\) when tuning the Fermi level close to the \(C_{2z}\mathcal{T}\)-protected Dirac cone.

The expressions in Eqs. \eqref{eqconductivity0} and \eqref{eqconductivity} for first-order and second-order longitudinal conductivity can be transformed into Fermi surface integrals by integration by parts, where \(\sigma^{(1)} \sim \int_{\mathcal{C}_{FS}} \partial_{k} \varepsilon_n\), \(\sigma^{(2)}_{\text{disp}} \sim \int_{\mathcal{C}_{FS}} \partial_{k}^2 \varepsilon_n\), and \(\sigma^{(2)}_{\text{geo}} \sim \int_{\mathcal{C}_{FS}} G_n\).

Due to the linear relation \( |\Delta \mu| \sim |\Delta k| \) near the Dirac point, where \( |\Delta \mu| \) is the Fermi level distance from the Dirac cone, we can approximate the elements in the above equations as follows: \(\int_{\mathcal{C}_{FS}} \sim 2\pi |\Delta k|\), \(\partial_{k_i} \sim |\Delta k|^{-1}\), \(\varepsilon_n \sim |\Delta k|\), and \(G_n^{aa} \sim |\Delta k|^{-3}\). Therefore, the resultant scaling behavior of linear and nonlinear longitudinal conductivity with respect to \( |\Delta \mu| \) would be:
\begin{equation}
\sigma^{(1)} \sim \tau |\Delta \mu|, \quad \frac{\partial \sigma^{(2)}_{\text{disp}}}{\partial B} \sim \tau^2, \quad \frac{\partial \sigma^{(2)}_{\text{geo}}}{\partial B} \sim \frac{1}{|\Delta \mu|^2}
\label{eq10}
\end{equation}

The corresponding EMCA coefficients from dispersion and quantum metric contributions then scale as:
\begin{equation}
\hat{\gamma}_{\text{disp}} \sim \frac{1}{|\Delta \mu|^2}, \quad \hat{\gamma}_{\text{geo}} \sim \frac{1}{\tau^2 |\Delta \mu|^4}
\label{eq11}
\end{equation}

From this simple scaling analysis, we deduce that the geometric part of EMCA, \(\hat{\gamma}_{\text{geo}}\), scales two orders faster than the traditional dispersion term, \(\hat{\gamma}_{\text{disp}}\), when tuning the Fermi level close to the \(C_{2z}\mathcal{T}\)-protected Dirac cone. Also, since both the dispersion \(\varepsilon\) and quantum metric \(G\) reverse sign when transitioning from the valence band to the conduction band, there are sign changes in both the dispersion and quantum metric parts of \(\frac{\partial \sigma_{ab;c}^{(2)}}{\partial B}\) and \(\hat{\gamma}\). These signatures will be shown in the following numerical calculations.

\subsection{Numerical Results}

We numerically calculated the linear and nonlinear conductivities, as well as the EMCA coefficient \(\hat{\gamma}\) of our chiral lattice model. The results are shown in Fig. \ref{fig2}.

Fig. \ref{fig2}(a) shows the linear longitudinal conductivity along the y-direction, \(\sigma^{y;y}\), at different Fermi levels \(\mu\). When the Fermi level is tuned close to the Dirac cones, the linear longitudinal conductivity \(\sigma_{y;y}\) behaves exactly as predicted in Eq. \eqref{eq10}, i.e., \(\sigma^{y;y} \sim |\Delta \mu|\).

Fig. \ref{fig2}(b) shows the dispersion and quantum metric contributions to \(\frac{\partial \sigma^{yy;y}}{\partial B}\), and Fig. \ref{fig2}(c) compares the dispersion and quantum metric contributions to the magnitude of the EMCA coefficient \(|\hat{\gamma}|\) at different Fermi levels \(\mu\) on a logarithmic scale, with the magnetic field applied in the y-direction. As predicted in Eqs. \eqref{eq10} and \eqref{eq11}, when the Fermi level is tuned close to the Dirac cones, \(\frac{\partial \sigma^{yy;y}_{\text{disp}}}{\partial B}\) scales as a constant with respect to \(|\Delta \mu|\), while \(\frac{\partial \sigma^{yy;y}_{\text{geo}}}{\partial B}\) scales as \(\frac{1}{|\Delta \mu|^2}\). Correspondingly, the quantum metric contribution to EMCA, \(\hat{\gamma}_{\text{geo}}\), scales faster than the dispersion contribution, \(\hat{\gamma}_{\text{disp}}\), by two orders of \(|\mu|\), resulting in a quantum metric EMCA-dominating regime near the Dirac cone, as indicated by the grey region in Fig. \ref{fig2}(c). This is due to the robustness of the \(C_{2z} \mathcal{T}\)-protected Dirac cones and the divergent quantum metric near the Dirac point.

From Fig. \ref{fig2}(b) and Fig. \ref{fig2}(c), we also observe that when the Fermi level is tuned away from the gapless point, the quantum geometric part decays rapidly, while the traditional dispersion part maintains its magnitude across the entire spectrum.

We note that there are discontinuous sign changes in both the dispersion and quantum metric parts of the EMCA when the Fermi level crosses the Dirac cone. This is caused by the sign differences of geometric quantities and velocities between the upper and lower Dirac bands. While the sign change in the dispersion term is small, due to the constant scaling of \(\frac{\partial \sigma_{\text{disp}}^{aa;a}}{\partial B} \sim |\Delta \mu|^0\) near the Dirac point, which is easily overwhelmed by subleading terms like \(\frac{\partial \sigma_{\text{disp}}^{aa;a}}{\partial B} \sim |\Delta \mu|^1\), the sign change in the quantum metric term is more pronounced due to its diverging scaling behavior as the Fermi level approaches the Dirac point.

We also calculated the transverse EMCA effect, where we keep the magnetic field in the $y$-direction while changing the electric field to the $x$-direction, measuring the Hall current in the $y$-direction. The nonlinear transverse conductivity also shows a dependence on both current and chirality, similar to the traditional longitudinal EMCA. As shown in Fig.~\ref{fig2} (d) and (e), the geometric component of the transverse EMCA exhibits a higher-order divergence that can easily surpass the conventional semiclassical term, which is solely due to band dispersion. Furthermore, the regime where the geometric transverse EMCA dominates is even broader than that of the longitudinal EMCA, potentially making it more accessible for experimental observation.

Although we employed a specific model, we emphasize that the geometric EMCA effect is a general feature in chiral materials with multiband structures, where the quantum metric is typically nonzero. While our model focuses on materials with a particular symmetry group, the geometric enhancement near Dirac cones or other geometric singularities in band structure like Weyl nodes is anticipated to be a universal characteristic in chiral materials. Our scaling analysis, therefore, is broadly applicable to any chiral or noncentrosymmetric materials with similar band structures.

Finally, we comment on the relation between our model calculations and real materials. In real chiral organic conductors, such as (DM-EDT-TTF)\(_2\)ClO\(_4\), the \(C_{2z} \mathcal{T}\)-protected Dirac cone still exists, but the Fermi level is near 0.18 eV, which is far from the Dirac point, as discussed in Appendix \ref{app0}. In principle, the dominant contribution to EMCA should still be the traditional dispersion part, \(\hat{\gamma}_{\text{disp}}\). However, even this dispersion contribution is smaller than the experimental results \cite{pop2014EMCA} by several orders of magnitude. We attribute this discrepancy to extrinsic effects, such as skew scattering and side-jump mechanisms, which occur in heavily disordered systems and are not accounted for in our intrinsic nonlinear response formalism. Nevertheless, we propose that a more pronounced geometric EMCA can be achieved experimentally by further hole-doping the material towards the Dirac points or finding other chiral materials with similar Dirac cone structure but a closer chemical potential.

\section{2D Topological Insulator}

\begin{figure}[tbp]
\includegraphics[width=0.99\linewidth]{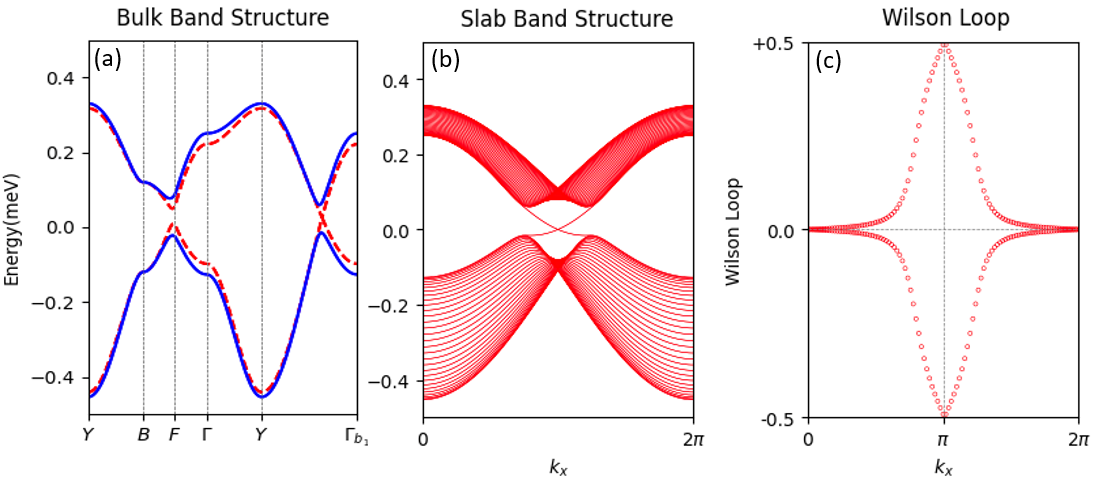}
\caption{ \label{FigS1} (a) The effect of N.N. SOC on the bulk band structure of the chiral lattice model. The red dashed line and blue solid line represent the model without and with N.N. SOC, respectively. The inclusion of SOC opens a gap at the Dirac point in the spinless model.
(b) Band structure of a finite slab of the chiral lattice model with SOC. A $C_{2z}$-breaking Rashba-type SOC is added, which breaks the degeneracy of the two sets of helical edge states.
(c) Wilson loop spectrum of the chiral lattice model with SOC. A nonzero winding number indicates a topological insulator state.}
\label{fig3}
\end{figure}

Additionally, introducing SOC to our chiral Dirac model leads to a 2D topological insulator phase, similar to graphene \cite{kane2005qsh}.
For a spinful model without external magnetic fields and SOC, the Bloch Hamiltonian is block-diagonal:
\begin{equation}
H_{tot}(\mathbf{k}) = H(\mathbf{k}) \otimes s_0 =  \begin{bmatrix}
H_\uparrow(\mathbf{k}) & \\
 & H_\downarrow(\mathbf{k}) \\
\end{bmatrix}
\end{equation}

Generally, we can write the $4 \times 4$ Bloch Hamiltonian as:
\begin{equation}
H_{tot}(\mathbf{k}) = \sum_{\pm} \sum_{i,j=0}^{4} d_{ij}^{(\pm)}(\mathbf{k}) s_i \sigma_j
\end{equation}
where the superscript \((\pm)\) represents the parity of \(d_{ij}(\mathbf{k})\), i.e., \(d_{ij}^{(\pm)}(\mathbf{k}) = \pm d_{ij}^{(\pm)}(-\mathbf{k})\). Similar to the spinless model, we can obtain the symmetry operators in the spinful model: \(C_{2z} = i \sigma_0 s_z\) and \(\mathcal{T} = -i \sigma_0 s_y \mathcal{K}\), which enforce certain \(d_{ij}^{(\pm)}\) terms to vanish. The nonzero elements are:
\begin{equation}
d_{ij}(\mathbf{k}) = \begin{bmatrix}
d_{00}^{(+)} & d_{01}^{(+)} & 0 & d_{03}^{(+)} \\
d_{10}^{(-)} & d_{11}^{(-)} & 0 & d_{13}^{(-)} \\
d_{20}^{(-)} & d_{21}^{(-)} & 0 & d_{23}^{(-)} \\
0 & 0 & d_{32}^{(+)} & 0 \\
\end{bmatrix}
\end{equation}

It is evident that our spinful model without SOC only consists of \(d_{00}^{(+)}\), \(d_{01}^{(+)}\), and \(d_{03}^{(+)}\). By adding SOC, we obtain general nonzero \(d_{ij}(\mathbf{k})\) terms that satisfy the symmetry constraints above. The \(d_{ij}\) dependence on \(\mathbf{k}\) depends on the explicit form of SOC added to the system. For example, with N.N. SOC:
\begin{equation}
H_{SOC} = \lambda_{SOC}\sum_{\langle i\alpha, j\beta \rangle} c_{i\alpha}^\dagger \mathbf{S}_{\alpha\beta} \cdot (\hat{\mathbf{d}}_{ij} \times \mathbf{e}_z) c_{j\beta}
\end{equation}
where \(S^i = \frac{s_i}{2}\), \(i = x, y, z\), is the i-th component of the spin operator. This induces a nonzero \(d_{32}^{(+)}(\mathbf{k}) = \lambda_{SOC} \cos(\mathbf{k} \cdot \mathbf{r}_{01})\) in the Bloch Hamiltonian:
\begin{equation}
\begin{aligned}
H_{tot}(\mathbf{k}) &= H(\mathbf{k}) \otimes s_0 + \lambda_{SOC} \cos(\mathbf{k} \cdot \mathbf{r}_{01}) \sigma_y \otimes s_z 
\end{aligned}
\end{equation}

The topological insulator state is confirmed by numerical calculation of the edge spectrum of a slab model and the Wilson loop. 
Fig. \ref{fig3}(a) shows that N.N. SOC gaps out the Dirac cone originally present in the spinless model. Fig. \ref{fig3}(b) shows the helical edge states inside the bulk gap in a slab model, indicating a conducting edge state. Fig. \ref{fig3}(c) shows the Wilson loop spectrum calculated along the \(k_x\) direction, where the nontrivial winding confirms the topological insulator state. Both these phenomena indicate a 2D topological insulator state.

We mention that for the present organic conductor, the spin-orbit coupling is negligible. Therefore, the topological insulator phase cannot appear in reality. However, for the completeness of our theory, it is still helpful to point out the connection between the Dirac cone and the topological insulator. In other materials with similar lattice structure but heavy elements, it is possible to realize the topological insulator.

\section{Summary}

In this work, we investigated EMCA in chiral crystals and uncovered a novel quantum geometric contribution to EMCA. We employed a 2D chiral lattice model and demonstrated that the quantum metric contribution can exceed the ordinary dispersion-related term in both traditionally longitudinal and a novel transverse EMCA when the Fermi level is tuned close to the Dirac point. The Dirac point is protected by \(C_{2z} \mathcal{T}\) symmetry, even in the presence of an external magnetic field, and can give rise to a 2D topological insulator state if strong SOC is present. Our results indicate that intriguing topological electronic structures can lead to significant EMCA in nonlinear magnetoresistance.

\begin{acknowledgements}

We thank helpful discussions with Geert Rikken, Yufei Zhao, Kamal Das and Tobias Holder. B.Y. acknowledges the financial support by the European Research Council (ERC Consolidator Grant ``NonlinearTopo'', No. 815869) and the ISF - Personal Research Grant (No. 2932/21) and the DFG (CRC 183, A02).

\end{acknowledgements}

\appendix

\begin{figure}[tbp]
\includegraphics[width=0.99\linewidth]{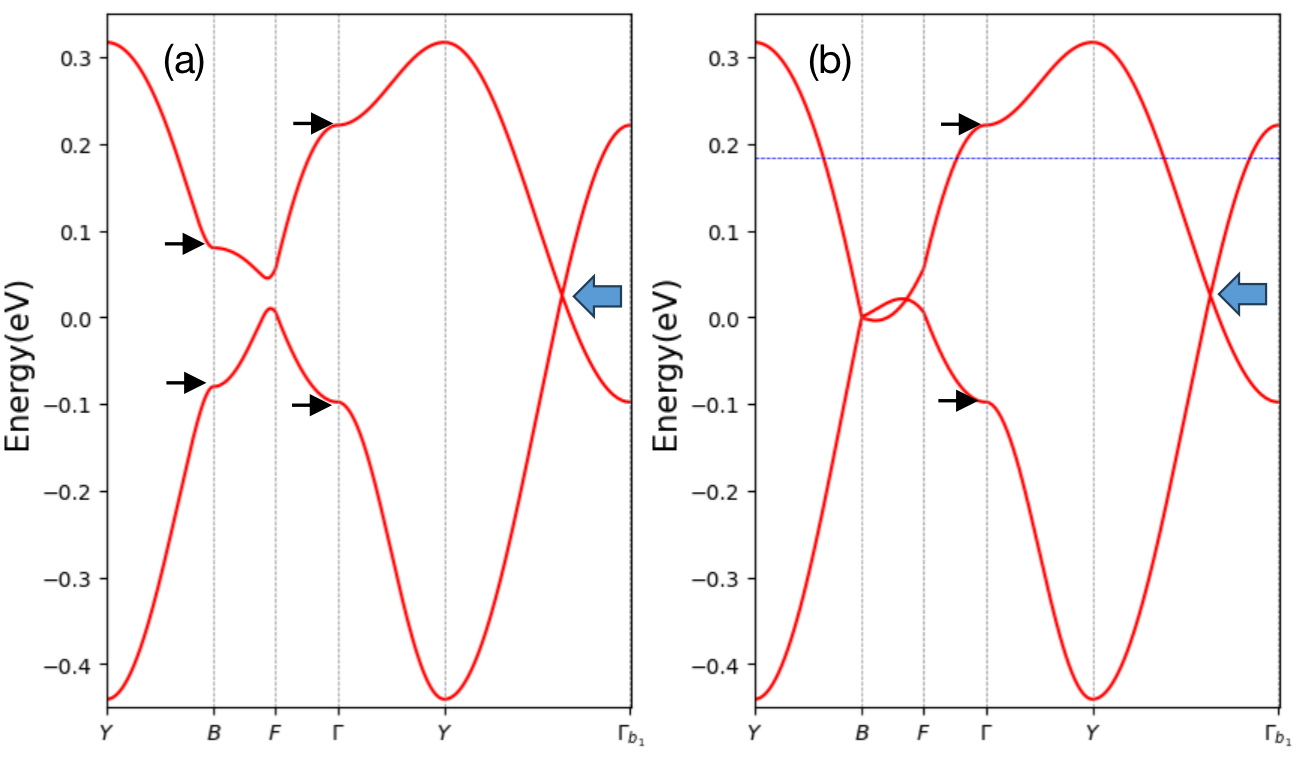}
\caption{Comparison of band structure and density of states (DOS) between the chiral lattice model and the real material. (a) The band structure and DOS of the chiral lattice model with parameters \(t_0 = 0.135\text{eV}\), \(t_1 = -0.055\text{eV}\), \(t_2 = 0.015\text{eV}\), \(\Delta t_2 = 0.02\text{eV}\), \(\delta U = \mu = 0\text{eV}\). (b) DFT-calculated band structure of one donor layer of the real chiral conductor (DM-EDT-TTF)${}_2$ClO${}_4$, consistent with the result in Ref.\cite{pop2014EMCA}, except that the Dirac point was overlooked. The blue and black arrows indicate Dirac points and van Hove singularities in the band structure. The blue dotted line represents the Fermi level of the real material.}
\label{appfig1}
\end{figure}

\section{Comparison of the model and real material}\label{app0}

The comparison of the band structure and density of states between the chiral lattice model presented in Fig. \ref{fig1} and the real material with DFT-calculated parameters is shown in Fig. \ref{appfig1}. The main differences lie in the anisotropic hopping parameter \(\Delta t_2\) and the Fermi level \(\mu\). In the real material, the \(C_{2z} \mathcal{T}\)-protected Dirac cones still exist, but they are hidden by the narrowed gap near the B-F line. In fact, the size of the gap near the B-F line is directly determined by the magnitude of the anisotropic hopping \(\Delta t_2\), i.e., the asymmetry or chirality of the model.

Additionally, the Fermi level is away from the Dirac cone, meaning that EMCA is still dominated by the dispersion contribution in the real material. Nevertheless, since the \(C_{2z} \mathcal{T}\)-protected Dirac cones persist, a strong geometric contribution to EMCA can still be expected if the Fermi level is properly tuned near the gapless points.

\section{Details on Peierls substitution}\label{appPeierls}

Assuming we have a Hamiltonian in the real-space basis,
\begin{equation}
  H = \sum_{i,j,\alpha,\beta} c_{i,\alpha}^\dagger h_{ij}^{\alpha \beta} c_{j,\beta}
\end{equation}
where \(i, j\) are unit cell indices and \(\alpha, \beta\) are orbital indices, the Bloch Hamiltonian has the expression,
\begin{equation}
    H^{\alpha\beta}(\mathbf{k}) = \sum_{i} e^{i \mathbf{k} \cdot \Delta \mathbf{r}} h_{ij}^{\alpha \beta}
\end{equation}
where \(\Delta\mathbf{r} \equiv \mathbf{R}_i + \mathbf{r}_{\alpha} - \mathbf{R}_j - \mathbf{r}_\beta\) is the hopping vector. Then, the original Bloch Hamiltonian can be formulated as:

\begin{equation}
    \begin{aligned}
        H^{00} (\mathbf{k}) & = (t_2 - \Delta t_2) e^{-i\mathbf{k} \cdot \mathbf{r}_1}  + (t_2 + \Delta t_2) e^{-i\mathbf{k} \cdot \mathbf{r}_2} + c.c. \\
        H^{11} (\mathbf{k}) & = (t_2 + \Delta t_2) e^{-i\mathbf{k} \cdot \mathbf{r}_1}  + (t_2 - \Delta t_2) e^{-i\mathbf{k} \cdot \mathbf{r}_2} + c.c. \\
        H^{01} (\mathbf{k}) & = H^{10} (\mathbf{k}) = t_0 e^{-i\mathbf{k} \cdot \mathbf{r}_{01}}  + t_1 e^{-i\mathbf{k} \cdot \mathbf{r}_{02}} + c.c. \\
    \end{aligned}
    \label{Hamiltonian_app}
\end{equation}

The Peierls substitution is as follows:
\begin{equation}
    H' = \sum_{i,j,\alpha, \beta} c_{i,\alpha}^\dagger h_{ij}^{\alpha \beta} e^{i \frac{e}{\hbar} \int_{\mathbf{R}_j + \mathbf{r}_\beta}^{\mathbf{R}_i + \mathbf{r}_{\alpha}} d\mathbf{r} \cdot \mathbf{A}(\mathbf{r})} c_{j,\beta}
\end{equation}
where \(\mathbf{A}\) is the vector potential that satisfies \(\nabla \times \mathbf{A} = \mathbf{B}\), and \(\mathbf{R}_i\) and \(\mathbf{r}_\alpha\) represent the position of unit cell \(i\) and the Wannier center of orbital \(\alpha\). The Peierls phase \(\frac{e}{\hbar} \int_{\mathbf{R}_j + \mathbf{r}_\beta}^{\mathbf{R}_i + \mathbf{r}_{\alpha}} d\mathbf{r} \cdot \mathbf{A}(\mathbf{r})\) calculates the magnetic flux through the area enclosed by the hopping direction and the coordinate axis.

In our model, we implemented a \(z\)-dependent vector potential \(\mathbf{A}\) to introduce a magnetic field \(\mathbf{B} = (B_x, B_y, 0)\) in the \(x\)-\(y\) plane.
\begin{equation}
    \mathbf{A}(\mathbf{r}) = (B_y z, -B_x z, 0)
\end{equation}

Thus, the Peierls phase is:
\begin{equation}
    \frac{e}{\hbar} \int_{\mathbf{R}_j + \mathbf{r}_\beta}^{\mathbf{R}_i + \mathbf{r}_{\alpha}} d\mathbf{r} \cdot \mathbf{A}(\mathbf{r}) = \frac{e\bar{z}}{\hbar} (\mathbf{B} \times \Delta\mathbf{r})_z
\end{equation}
where \(\bar{z} \equiv (\mathbf{R}_i + \mathbf{r}_{\alpha} + \mathbf{R}_j + \mathbf{r}_\beta)_z\) represents the \(z\)-component of the average position. In our model, we choose the origin of the \(z\)-coordinate at the center of the two enantiomer layers, so the interlayer hopping vectors \(\mathbf{r}_{01}\) and \(\mathbf{r}_{02}\) do not gain Peierls phases, as \(\bar{z} = 0\). Thus, the Peierls-substituted Bloch Hamiltonian is:
\begin{widetext}
    \begin{equation}
    \begin{aligned}
        H^{00} (\mathbf{k};\mathbf{B}) & = (t_2 - \Delta t_2) e^{-i\mathbf{k} \cdot \mathbf{r}_1} e^{-i \frac{e \Delta z}{\hbar} (\mathbf{B} \times \mathbf{r}_1)_z}  + (t_2 + \Delta t_2) e^{-i\mathbf{k} \cdot \mathbf{r}_2} e^{-i \frac{e \Delta z}{\hbar} (\mathbf{B} \times \mathbf{r}_2)_z} + c.c. \\
        H^{11} (\mathbf{k};\mathbf{B}) & = (t_2 + \Delta t_2) e^{-i\mathbf{k} \cdot \mathbf{r}_1} e^{-i \frac{e (-\Delta z)}{\hbar} (\mathbf{B} \times \mathbf{r}_1)_z} + (t_2 - \Delta t_2) e^{-i\mathbf{k} \cdot \mathbf{r}_2} e^{-i \frac{e (-\Delta z)}{\hbar} (\mathbf{B} \times \mathbf{r}_2)_z} + c.c. \\
        H^{01} (\mathbf{k};\mathbf{B}) & = H^{10} (\mathbf{k}) = t_0 e^{-i\mathbf{k} \cdot \mathbf{r}_{01}}  + t_1 e^{-i\mathbf{k} \cdot \mathbf{r}_{02}} + c.c. \\
    \end{aligned}
    \label{Hamiltonian_app_adding_B}
\end{equation}
\end{widetext}

After rearrangement, Eqs.~\ref{Hamiltonian_app} and \ref{Hamiltonian_app_adding_B} become Eqs.~\ref{Hamiltonian} and \ref{Hamiltonian_adding_B} as shown in the main text.

\section{Theory of EMCA}\label{app1}

\subsection{Phenomenology}

For chiral materials in a magnetic field, both theoretical proposals and experimental measurements give:

\begin{equation}
R^{\chi}(\mathbf{I}; \mathbf{B}) = R_0 (1 + \beta B^2 + \gamma^{\chi} \mathbf{B} \cdot \mathbf{I})
\end{equation}

Here, \(R\) represents the resistance of a certain chiral material, and \(\chi = \pm 1\) symbolizes the left/right-handedness of the chiral material. The \(\mu^2 B^2\) term is the diamagnetic part, and \(\gamma^{D/L} \mathbf{B} \cdot \mathbf{I}\) is the EMCA part.

Since left- and right-handedness are related by spatial inversion (plus certain rotations), by symmetry analysis we have:
\begin{equation}
\gamma^\chi = - \gamma^{-\chi}
\end{equation}

Now we discuss the units used here, as the terms inside the parentheses are dimensionless. In SI units, \(\mathbf{B}\) is in Tesla (T) and \(\mathbf{I}\) is in Amperes (A). The unit for \(\gamma^{\chi}\) is therefore \(T^{-1}A^{-1}\).

The discussion above relates to EMCA in terms of resistance, but we can convert it to conductivity, which is more convenient when discussing its relation to the nonlinear response later. We also use current density \(\mathbf{j}\) instead of current \(\mathbf{I}\) for the same reason.

For longitudinal resistivity:
\begin{equation}
\rho_{a;a}(\mathbf{j}; \mathbf{B}) = \rho_0 (1 + \mu^2 B^2 + \hat{\gamma}^{D/L} \mathbf{B} \cdot \mathbf{j})
\end{equation}

Neglecting transverse resistivity and assuming \(\mu^2 B^2 \ll 1\) and \(\hat{\gamma}^{D/L} \mathbf{B} \cdot \mathbf{j} \ll 1\), we have a similar EMCA expression for longitudinal conductivity:
\begin{equation}
\sigma_{a;a} (\mathbf{j}; \mathbf{B}) = \sigma_0 (1 - \mu^2 B^2 - \hat{\gamma}^{D/L} \mathbf{B} \cdot \mathbf{j})
\end{equation}

Since the SI unit for current density is \(A \cdot m^{-2}\), the unit for the corresponding EMCA coefficient \(\hat{\gamma}\) is \(m^2 T^{-1} A^{-1}\).

\subsection{EMCA as a nonlinear response}

We typically consider the current density response \(\mathbf{j}\) up to different orders of the electric field \(\mathbf{E}\):
\begin{equation}
j_a = \sigma_{b;a}^{(1)} E_b + \frac{1}{2!} \sigma_{bc;a}^{(2)} E_b E_c + \frac{1}{3!} \sigma_{bcd;a}^{(3)} E_b E_c E_d + \dots
\end{equation}
Here, \(\sigma^{(n)}\) is the \(n\)-th order conductivity, and the Einstein summation convention is implied. We can assume an inverse relation with \(\rho^{(n)}\), the \(n\)-th order resistivity:
\begin{equation}
E_a = \rho_{b;a}^{(1)} j_b + \frac{1}{2!} \rho_{bc;a}^{(2)} j_b j_c + \frac{1}{3!} \rho_{bcd;a}^{(3)} j_b j_c j_d + \dots
\end{equation}
Although the relations between arbitrary orders of \(\rho^{(m)}\) and \(\sigma^{(n)}\) are generally complex, to the first order it's simply the inversion of the matrix:
\begin{equation}
(\sigma^{(1)})^{-1}_{a;b} = \rho_{a;b}^{(1)}
\end{equation}

For 2D materials:
\begin{equation}
\begin{aligned}
\rho^{(1)} & = \begin{bmatrix}
\rho^{(1)}_{x;x} & \rho^{(1)}_{x;y} \\
\rho^{(1)}_{y;x} & \rho^{(1)}_{y;y} \\
\end{bmatrix} = (\sigma^{(1)})^{-1} \\
& = \frac{1}{\sigma^{(1)}_{x;x} \sigma^{(1)}_{y;y} - \sigma^{(1)}_{x;y} \sigma^{(1)}_{y;x}} \begin{bmatrix}
\sigma^{(1)}_{y;y} & -\sigma^{(1)}_{x;y} \\
-\sigma^{(1)}_{y;x} & \sigma^{(1)}_{x;x} \\
\end{bmatrix}
\end{aligned}
\end{equation}
\begin{equation}
\begin{aligned}
\sigma^{(1)} & = \begin{bmatrix}
\sigma^{(1)}_{x;x} & \sigma^{(1)}_{x;y} \\
\sigma^{(1)}_{y;x} & \sigma^{(1)}_{y;y} \\
\end{bmatrix} = (\rho^{(1)})^{-1} \\
& = \frac{1}{\rho^{(1)}_{x;x} \rho^{(1)}_{y;y} - \rho^{(1)}_{x;y} \rho^{(1)}_{y;x}} \begin{bmatrix}
\rho^{(1)}_{y;y} & -\rho^{(1)}_{x;y} \\
-\rho^{(1)}_{y;x} & \rho^{(1)}_{x;x} \\
\end{bmatrix}
\end{aligned}
\end{equation}

If the first-order transverse conductivity/resistivity is negligible compared to the longitudinal conductance/resistivity, we can set \(\sigma^{(1)}_{x;y} = \sigma^{(1)}_{y;x} = \rho^{(1)}_{x;y} = \rho^{(1)}_{y;x} = 0\), and we have:
\begin{equation}
\sigma^{(1)} = \begin{bmatrix}
\sigma^{(1)}_{x;x} & 0 \\
0 & \sigma^{(1)}_{y;y} \\
\end{bmatrix} = \begin{bmatrix}
1/\rho^{(1)}_{x;x} & 0 \\
0 & 1/\rho^{(1)}_{y;y} \\
\end{bmatrix}
\end{equation}

Now, to get EMCA from these formulas, the total conductivity \(\sigma_{a;b}\) can be defined as:
\begin{equation}
\sigma_{b;a} \equiv \frac{\partial j_a}{\partial E_b}
\end{equation}

Then, the expression for conductivity in terms of the electric field $\mathbf{E}$ can be written as,
\begin{equation}
\sigma_{b;a}(\mathbf{E}) = \sigma^{(1)}_{b;a} + \sigma^{(2)}_{c_1 b;a} E_{c_1} + \frac{1}{2!} \sigma^{(3)}_{c_1 c_2 b;a} E_{c_1} E_{c_2} + \dots
\end{equation}

We further expand \(E_{c_i}\) in terms of the \(n\)-th order resistivity \(\rho^{(n)}\) and current density \(j_{d}\):
\begin{equation}
E_{c_i} = \rho_{d_1;c_i}^{(1)} j_{d_1} + \frac{1}{2!} \rho_{d_1 d_2;c_i}^{(2)} j_{d_1} j_{d_2} + \frac{1}{3!} \rho_{d_1 d_2 d_3;c_i}^{(3)} j_{d_1} j_{d_2} j_{d_3} + \dots
\end{equation}

Then, the expression for conductivity in terms of the current density $\mathbf{j}$ can be written as,
\begin{equation}
\begin{aligned}
\sigma_{b;a}(\mathbf{j}) & = \sigma^{(1)}_{b;a} \\
& + \sigma^{(2)}_{c_1 b;a} (\rho_{d_1;c_1}^{(1)} j_{d_1} + \rho_{d_1 d_2;c_1}^{(2)} j_{d_1} j_{d_2} + \dots) \\
& + \dots
\end{aligned}
\end{equation}

Furthermore, the system has another parameter, the magnetic field \(\mathbf{B}\). We assume it's a system parameter, and each order of conductivity/resistivity has a certain dependence on it, allowing a Taylor expansion:
\begin{equation}
\begin{aligned}
\sigma^{(m)} 
& \equiv (\sigma^{(m)})_0 + (\partial_{k_1} \sigma^{(m)})_0 B_{k_1} \\
& + \frac{1}{2!} (\partial_{k_1} \partial_{k_2} \sigma^{(m)})_0 B_{k_1} B_{k_2} + \dots \\
\rho^{(m)} & \equiv (\rho^{(m)})_0 + (\partial_{k_1} \rho^{(m)})_0 B_{k_1} \\
& + \frac{1}{2!} (\partial_{k_1} \partial_{k_2} \rho^{(m)})_0 B_{k_1} B_{k_2} + \dots
\end{aligned}
\end{equation}

So, the expression for conductivity in terms of the current density $\mathbf{j}$ in an external magnetic field $\mathbf{B}$ can be written as,
\begin{widetext}
\begin{equation}
\begin{aligned}
\sigma_{b;a}(\mathbf{j}; \mathbf{B}) & = \sigma^{(1)}_{b;a} + \sigma^{(2)}_{c_1 b;a} (\rho_{d_1;c_1}^{(1)} j_{d_1} + \frac{1}{2!} \rho_{d_1 d_2;c_1}^{(2)} j_{d_1} j_{d_2} + \dots) + \dots \\
& = (\sigma^{(1)}_{b;a})_0 + (\partial_{k_1} \sigma^{(1)}_{b;a})_0 B_{k_1} + \frac{1}{2!} (\partial_{k_1} \partial_{k_2} \sigma^{(1)}_{b;a})_0 B_{k_1} B_{k_2} + \dots \\
& + \left[ (\sigma^{(2)}_{c_1 b;a})_0 + (\partial_{k_1} \sigma^{(2)}_{c_1 b;a})_0 B_{k_1} + \frac{1}{2!} (\partial_{k_1} \partial_{k_2} \sigma^{(2)}_{a;c_1 b})_0 B_{k_1} B_{k_2} + \dots \right] \\
& \times \left[ \left( (\rho^{(1)}_{d_1;c_1})_0 + (\partial_{k_1} \rho^{(1)}_{d_1;c_1})_0 B_{k_1} + \frac{1}{2!} (\partial_{k_1} \partial_{k_2} \rho^{(1)}_{d_1;c_1})_0 B_{k_1} B_{k_2} + \dots \right) j_{d_1} \right. \\
& \quad \left. + \frac{1}{2!} \left( (\rho^{(2)}_{d_1 d_2;c_1})_0 + (\partial_{k_1} \rho^{(2)}_{d_1 d_2;c_1})_0 B_{k_1} + \frac{1}{2!} (\partial_{k_1} \partial_{k_2} \rho^{(2)}_{d_1 d_2;c_1})_0 B_{k_1} B_{k_2} + \dots \right) j_{d_1} j_{d_2} + \dots \right] + \dots \\
& = (\sigma^{(1)}_{b;a})_0 + (\partial_{k_1} \sigma^{(1)}_{b;a})_0 B_{k_1} + (\sigma^{(2)}_{c_1 b;a})_0 (\rho^{(1)}_{d_1;c_1})_0 j_{d_1} + \frac{1}{2!} (\partial_{k_1} \partial_{k_2} \sigma^{(1)}_{b;a})_0 B_{k_1} B_{k_2} \\
& + (\sigma^{(2)}_{c_1 b;a})_0 (\partial_{k_1} \rho^{(1)}_{d_1;c_1})_0 B_{k_1} j_{d_1} + (\partial_{k_1} \sigma^{(2)}_{c_1 b;a})_0 (\rho^{(1)}_{d_1;c_1})_0 B_{k_1} j_{d_1} \\
& + \frac{1}{2!} (\sigma^{(2)}_{c_1 b;a})_0 (\rho^{(2)}_{d_1 d_2;c_1})_0 j_{d_1} j_{d_2} + \frac{1}{2!} (\sigma^{(3)}_{c_1 c_2 b;a})_0 (\rho^{(1)}_{d_1;c_1})_0 (\rho^{(1)}_{d_2;c_2})_0 j_{d_1} j_{d_2} \\
& + \mathcal{O}(j^3, j^2B, jB^2, B^3)
\end{aligned}
\end{equation}

From symmetry analysis, we have:
\begin{equation}
(\partial_{k_1} \sigma^{(1)}_{b;a})_0 = 0 
\quad \Leftrightarrow \quad  (\partial_{k_1} \rho^{(1)}_{d_1;c_1})_0 = 0 ,\, (\sigma^{(2)}_{c_1 b;a})_0 = 0
\end{equation}

Then, the conductivity formula to the second order of current density $\mathbf{j}$ and magnetic field $\mathbf{B}$ is:
\begin{equation}
\begin{aligned}
\sigma_{b;a}(\mathbf{j}; \mathbf{B}) & = \underbrace{(\sigma^{(1)}_{b;a})_0}_{\sigma_0} + \underbrace{\frac{1}{2!} (\partial_{k_1} \partial_{k_2} \sigma^{(1)}_{b;a})_0 B_{k_1} B_{k_2}}_{-\sigma_0 \mu^2 B^2} + \underbrace{(\partial_{k_1} \sigma^{(2)}_{a;c_1 b})_0 (\rho^{(1)}_{d_1;c_1})_0 B_{k_1} j_{d_1}}_{-\sigma_0 \gamma^{D/L} \mathbf{j} \cdot \mathbf{B}} \\
& + \frac{1}{2!} (\sigma^{(3)}_{c_1 c_2 b;a})_0 (\rho^{(1)}_{d_1;c_1})_0 (\rho^{(1)}_{d_2;c_2})_0 j_{d_1} j_{d_2} + \mathcal{O}(j^3, j^2B, jB^2, B^3)
\end{aligned}
\end{equation}
\end{widetext}

By neglecting the transverse current response and assuming isotropy in the material, we have:
\begin{equation}
\begin{aligned}
& \sigma_{a;a}(\mathbf{j};\mathbf{B}) \\
& = (\sigma^{(1)}_{a;a})_0 + \frac{1}{2!} (\frac{\partial^2 \sigma^{(1)}_{a;a}}{\partial B^2})_0 B^2 + (\frac{\partial \sigma^{(2)}_{aa;a}}{\partial B})_0 \frac{1}{(\sigma^{(1)}_{a;a})_0} \mathbf{B} \cdot \mathbf{j} + \dots \\
& = (\sigma^{(1)}_{a;a})_0 \left( 1 + \frac{1}{2!} \frac{(\frac{\partial^2 \sigma^{(1)}_{a;a}}{\partial B^2})_0}{(\sigma^{(1)}_{a;a})_0} B^2 + \frac{(\frac{\partial \sigma^{(2)}_{aa;a}}{\partial B})_0}{{(\sigma^{(1)}_{a;a})_0}^2} \mathbf{B} \cdot \mathbf{j} \right) \\
& \equiv \sigma_0 (1 - \mu^2 B^2 - \hat{\gamma}^{D/L} \mathbf{B} \cdot \mathbf{j})
\end{aligned}
\end{equation}

Thus, the EMCA coefficient, defined from the general nonlinear response function, would be:
\begin{equation}
\hat{\gamma} = - \frac{\left( \frac{\partial \sigma^{(2)}_{aa;a}}{\partial B_a} \right)_0}{(\sigma^{(1)}_{a;a})_0^2} = - \lim_{B \to 0} \frac{\sigma^{(2)}_{aa;a}(B_a)}{B_a \left[ \sigma^{(1)}_{a;a}(B_a) \right]^2}
\end{equation}

Therefore, EMCA is related to the first- and second-order current responses of chiral materials, and we can calculate the EMCA coefficient by calculating the first- and second-order conductivities and their dependence on the external magnetic field.

\section{Derivation of Nonlinear Conductivity}\label{app2}

\subsubsection{Semiclassical nonlinear conductivity}

Semiclassically, the current density of the system is,
\begin{equation}
\mathbf{j} = - e \sum_{n} \int \frac{d^D \mathbf{k}}{(2\pi)^D} f_n \mathbf{v}_n
\end{equation}

Here, \(f_n\) is the distribution function of the \(n\)-th band, \(-e\) is the charge of the electron, \(\mathbf{v}_n\) is the velocity of the \(n\)-th band, and the external magnetic field \(\mathbf{B}\) is treated as a model parameter, while the external electric field \(\mathbf{E}\) is considered a perturbation to the system.

Semiclassically, the distribution function \(f(\mathbf{r}, \mathbf{k}, t)\) satisfies the Boltzmann transport equation, which is,
\begin{equation}
\partial_t f + \frac{\mathbf{F}}{\hbar} \cdot \nabla_\mathbf{k} f + \mathbf{v} \nabla_\mathbf{r} f = \mathcal{I}\left[f\right]
\end{equation}
where \(\mathbf{F}\) is the semiclassical force in the presence of an electric field,
\begin{equation}
\mathbf{F} = - e \mathbf{E}
\end{equation}

\(\mathcal{I}\left[f\right]\) is the collision integral. After implementing the constant relaxation time approximation, it yields
\begin{equation}
\mathcal{I}\left[f\right] = - \frac{f - f^{(0)}}{\tau}
\end{equation}
where \(\tau\) is the scattering time and \(f^{(0)}\) is the Fermi-Dirac distribution.
Assuming translational invariance of the system, we can neglect the \(\nabla_\mathbf{r} f\) term, and after Fourier transformation, we get,
\begin{equation}
i \omega \tilde{f} + \frac{-e \mathbf{E}}{\hbar} \cdot \nabla_{\mathbf{k}} \tilde{f} = -\frac{\tilde{f}-\tilde{f}^{(0)}}{\tau}
\end{equation}

Perturbatively expanding \(f\) order by order yields,
\begin{equation}
f(t) = f^{(0)}(t) + \sum_{l=1}^\infty f^{(l)}(t)
\end{equation}
where \(f^{(l)}(t) = \mathcal{O}(\mathbf{E}^l)\), and we have a similar Fourier transformed version,
\begin{equation}
\tilde{f}(\omega) = \tilde{f}_0(\omega) + \sum_{l=1}^\infty \tilde{f}^{(l)}(\omega)
\end{equation}

Then \(\tilde{f}_n^{(l)}\) can be obtained recursively,
\begin{equation}
\tilde{f}_n^{(l)} = \frac{\frac{e}{\hbar} \mathbf{E}^l \nabla_{\mathbf{k}} \tilde{f}_n^{(l-1)}}{i \omega+1 / \tau} = \left(\frac{e / \hbar}{i \omega+1 / \tau}\right)^l \mathbf{E}^l \nabla_{\mathbf{k}}^l \tilde{f}_n^{(0)}
\end{equation}
where \(\mathbf{E}^l \nabla_{\mathbf{k}} = E^{a_1} E^{a_2} \ldots E^{a_l} \partial_{k^{a_1}} \partial_{k^{a_2}} \ldots \partial_{k^{a_l}}\).
In the direct current limit, where \(\omega \ll 1 / \tau\),
\begin{equation}
\lim _{\omega \rightarrow 0} \tilde{f}_n^{(l)} \approx\left(\frac{e \tau}{\hbar}\right)^l \mathbf{E}^l \nabla_{\mathbf{k}}^l \tilde{f}_n^{(0)}
\end{equation}
or equivalently,
\begin{equation}
\lim _{\omega \rightarrow 0} f_n^{(l)} \approx\left(\frac{e \tau}{\hbar}\right)^l \mathbf{E}^l \nabla_{\mathbf{k}}^l f_n^{(0)}
\end{equation}

Next, we determine the expansion of \(\mathbf{v}_n\). The group velocity has contributions from the band dispersion \(\varepsilon_n\) and the Berry curvature \(\mathbf{\Omega}_n\), as follows:
\begin{equation}
\begin{aligned}
\mathbf{v}_n^{(0)} & =\frac{1}{\hbar} \frac{\partial \varepsilon_n^{(0)}}{\partial \mathbf{k}} \\
\mathbf{v}_n^{(1)} & =\frac{1}{\hbar} \frac{\partial \varepsilon_n^{(1)}}{\partial \mathbf{k}}-\frac{e}{\hbar} \mathbf{E} \times \boldsymbol{\Omega}_n^{(0)} \\
\mathbf{v}_n^{(2)} & =\frac{1}{\hbar} \frac{\partial \varepsilon_n^{(2)}}{\partial \mathbf{k}}-\frac{e}{\hbar} \mathbf{E} \times \boldsymbol{\Omega}_n^{(1)} \\
\mathbf{v}_n^{(3)} & =\frac{1}{\hbar} \frac{\partial \varepsilon_n^{(3)}}{\partial \mathbf{k}}-\frac{e}{\hbar} \mathbf{E} \times \boldsymbol{\Omega}_n^{(2)}
\end{aligned}
\end{equation}

Here, we also enforce an expansion in orders of \(\mathbf{E}\):
\begin{equation}
\mathbf{v}_n^{(n)} = \mathcal{O}(\mathbf{E}^n)
,\,
\varepsilon_n^{(n)} = \mathcal{O}(\mathbf{E}^n)
,\,
\mathbf{\Omega}_n^{(n)} = \mathcal{O}(\mathbf{E}^n)
\end{equation}
which yields the \(l\)-th order electron current for \(l = 0, 1, 2, 3\):
\begin{equation}
\begin{aligned}
& \mathbf{j}^{(0)}=-e \int_k \sum_n f_n^{(0)} \mathbf{v}_n^{(0)}=0 \\
& \mathbf{j}^{(1)}=-e \int_k \sum_n f_n^{(1)} \mathbf{v}_n^{(0)}+f_n^{(0)} \mathbf{v}_n^{(1)} \\
& \mathbf{j}^{(2)}=-e \int_k \sum_n f_n^{(2)} \mathbf{v}_n^{(0)}+f_n^{(1)} \mathbf{v}_n^{(1)}+f_n^{(0)} \mathbf{v}_n^{(2)} \\
\end{aligned}
\end{equation}

Comparing with the definition for \(n\)-th order conductivity:
\begin{equation}
j_a = \sigma_{a;b}^{(1)} E_b + \frac{1}{2!} \sigma_{a;bc}^{(2)} E_b E_c + \frac{1}{3!} \sigma_{a;bcd}^{(3)} E_b E_c E_d + \dots
\end{equation}
we have:
\begin{equation}
\sigma_{a;b_1 \dots b_n}^{(n)} = \left. \frac{\partial^n j_a}{\partial E_{b_1} \dots \partial E_{b_n}} \right|_{\mathbf{E} = 0}
\end{equation}

Therefore, if we want to calculate up to 2nd order current response, we should calculate:
\begin{equation}
f^{(0)},\,f^{(1)},\,f^{(2)}
\end{equation}
\begin{equation}
\varepsilon^{(0)},\,\varepsilon^{(1)},\,\varepsilon^{(2)}
\end{equation}
\begin{equation}
\mathbf{\Omega}^{(0)},\,\mathbf{\Omega}^{(1)}
\end{equation}

As shown above, we can calculate \(f^{(n)}\) by \(\lim _{\omega \rightarrow 0} f_n^{(l)} \approx\left(\frac{e \tau}{\hbar}\right)^l \mathbf{E}^l \nabla_{\mathbf{k}}^l f_n^{(0)}\), so now we need to calculate the band structure and Berry curvature after applying electromagnetic perturbation, i.e., we need to implement quantum perturbation theory to calculate the eigenvalues and eigenstates.

\subsection{Perturbation Theory and Schrieffer–Wolff transformation}

\subsubsection{General Form of SW Transformation}

As mentioned above, we need to calculate eigenvalues up to the 3rd order perturbative correction, i.e., \(\mathcal{O}(\mathbf{E}^3)\), and wavefunction up to the 2nd order, i.e., \(\mathcal{O}(\mathbf{E}^2)\).

Usually, we are familiar with perturbative expansion of eigenvalues up to 2nd order and that of eigenstates up to 1st order. To avoid a painstaking derivation of higher-order perturbation theory, we implement a Schrieffer-Wolff (SW) transformation, which changes the wave function basis here, transforming the lowest order perturbation to the same order of \(\mathcal{O}(\mathbf{E}^2)\), i.e.,
\begin{equation}
H \rightarrow H' \equiv e^{S} H e^{-S},\,|n\rangle \rightarrow |n'\rangle \equiv e^S |n\rangle 
\end{equation}
\begin{equation}
H' = H_0' + H_{1}'
\end{equation}
\begin{equation}
H_{1}' = \mathcal{O}(\mathbf{E}^2)
\end{equation}

Now, the perturbative expansions of the eigenvalue and eigenstate only require the 2nd order eigenvalue and 1st order eigenstate perturbation theory for \(H'\) to cover \(\mathcal{O}(\mathbf{E}^3)\) for the former and \(\mathcal{O}(\mathbf{E}^2)\) for the latter. Throughout the discussion below, the prime \('\) represents the eigenstates and operators after the SW transformation.

We explicitly implement the SW transformation. Before doing that, we separate the original Hamiltonian \(H\) into diagonal and off-diagonal parts:
\begin{equation}
H = H^{(0)} + e \mathbf{E} \cdot \mathbf{r} \equiv H_0 + H_1
\end{equation}
where \(H^{(0)}\) is the Hamiltonian up to \(\mathcal{O}(\mathbf{E}^0)\), i.e., the original lattice Hamiltonian, and \(H_0\), \(H_1\) correspond to the diagonal and off-diagonal parts of the full Hamiltonian after applying the external field, which reads:
\begin{equation}
\begin{aligned}
H_0 & \equiv H^{(0)} + \sum_{\mathbf{k}} \sum_{n} \left( e \mathbf{E} \cdot \langle \psi_{n\mathbf{k}}^{(0)} | \mathbf{r} | \psi_{n\mathbf{k}}^{(0)} \rangle \right) | \psi_{n\mathbf{k}}^{(0)} \rangle \langle \psi_{n\mathbf{k}}^{(0)} | \\
& \equiv \sum_{\mathbf{k}} \sum_{n} \left( \varepsilon_n^{(0)} (\mathbf{k}) + e \mathbf{E} \cdot \mathbf{A}_{nn}^{(0)}(\mathbf{k}) \right) | \psi_{n\mathbf{k}}^{(0)} \rangle \langle \psi_{n\mathbf{k}}^{(0)} |
\end{aligned}
\end{equation}
\begin{equation}
\begin{aligned}
H_1 & \equiv \sum_{\mathbf{k} \neq \mathbf{k}'} \sum_{n \neq m} \left( e \mathbf{E} \cdot \langle \psi_{n\mathbf{k}}^{(0)} | \mathbf{r} | \psi_{m\mathbf{k}'}^{(0)} \rangle \right) | \psi_{n\mathbf{k}}^{(0)} \rangle \langle \psi_{m\mathbf{k}'}^{(0)} | \\
& =  \sum_{\mathbf{k}} \sum_{n \neq m} \left( e \mathbf{E} \cdot \langle \psi_{n\mathbf{k}}^{(0)}  | \mathbf{r} | \psi_{m\mathbf{k}}^{(0)}  \rangle \right) | \psi_{n\mathbf{k}}^{(0)}  \rangle \langle \psi_{m\mathbf{k}}^{(0)}  | \\
& \equiv \sum_{\mathbf{k}} \sum_{n \neq m} \left( e \mathbf{E} \cdot \mathbf{A}_{nm}^{(0)} (\mathbf{k}) \right) | \psi_{n\mathbf{k}}^{(0)}  \rangle \langle \psi_{m\mathbf{k}}^{(0)}  |
\end{aligned}
\end{equation}

Here, \(\varepsilon_n^{(n)}(\mathbf{k})\), \(|\psi_{n\mathbf{k}}^{(0)}\rangle\) represents the eigenvalues and eigenstates (Bloch wavefunction) of the \(n\)-th band at crystal momentum \(\mathbf{k}\) of the original Hamiltonian \(H^{(0)}\) before applying the electric field, and \(\mathbf{A}_{nm}^{(0)}(\mathbf{k})\) symbolizes the (0th order) Berry connection.

We emphasize that the upper index \((\dots)^{(n)}\) indicates considering \(e \mathbf{E} \cdot \mathbf{r}\) as the perturbation, since we only know about the dispersion \(\varepsilon_n^{(0)}(\mathbf{k})\) and the eigenstate (Bloch wavefunction) \(|\psi_{n\mathcal{k}}^{(0)} \rangle\), so we have to transform all the quantities we want to calculate into forms of \((\dots)^{(n)}\).

We can also write the SW transformation as:
\begin{equation}
H' = e^{\epsilon S} H e^{-\epsilon S}
\end{equation}
and we regard the perturbative parameter \(\epsilon\) as the marker for the order of the original perturbation theory, i.e., we set \(\epsilon\) as a quantity proportional to the electric field amplitude \(|\mathbf{E}|\),
\begin{equation}
\epsilon \propto |\mathbf{E}|
\end{equation}

Accordingly, we can write the separation of the diagonal and off-diagonal parts of \(H\) as:
\begin{equation}
H = H_0 + \epsilon H_1
\end{equation}
Then we can check the correspondence of the order of the perturbation theory we are doing both before and after the SW transformation by expanding the SW transformation order by order with respect to the parameter \(\epsilon\):
\begin{equation}
\begin{aligned}
H' & = \sum_{n = 0}^{\infty} \frac{\epsilon^n}{n!} \underbrace{[S,[S,\dots,[S,H]\dots]]}_{n\text{ nested commutators}} \\
& = H + \epsilon [S, H] + \frac{\epsilon^2}{2!}[S,[S,H]] + \frac{\epsilon^3}{3!} [S,[S,[S,H]]] + \dots \\
& = H_0 + \epsilon \left( H_1 + [S,H_0] \right) + \epsilon^2 \left( [S,H_1] + \frac{1}{2} [S,[S,H_0]] \right) \\
& + \dots
\end{aligned}
\end{equation}

Now we're going to specify the operator \(S\), which should be anti-Hermitian to guarantee the SW transformation is a unitary transformation.

Secondly, the transformation should ensure that the SW-transformed Hamiltonian has a perturbation term only up to \(\mathcal{O}(\mathbf{E}^2) \sim \mathcal{O}(\epsilon^2)\), then we have to set:
\begin{equation}
H_1 + [S,H_0] = 0
\end{equation}

By assuming \(S\) has the same discrete translational symmetry as \(H^{(0)}\), we can write the operator \(S\) in the Bloch basis, which is block diagonal in momentum space:
\begin{equation}
\begin{aligned}
S & = \sum_{\mathbf{k}} \sum_{n,m} \left( \langle \psi_{n\mathbf{k}}^{(0)} | S | \psi_{m\mathbf{k}}^{(0)} \rangle \right) | \psi_{n\mathbf{k}}^{(0)} \rangle \langle \psi_{m\mathbf{k}}^{(0)} | \\
& = \sum_{\mathbf{k}} \sum_{n,m} S_{nm} | \psi_{n\mathbf{k}}^{(0)} \rangle \langle \psi_{m\mathbf{k}}^{(0)} |
\end{aligned}
\end{equation}

Here, we define the matrix element \(S_{nm}\equiv\langle \psi_{n\mathbf{k}}^{(0)} | S | \psi_{m\mathbf{k}}^{(0)} \rangle\), and we omit the \((\mathbf{k})\) here as well. We can also write this constraint in the basis of \(| \psi_{n \mathbf{k}}^{(0)} \rangle\) as:
\begin{equation}
\begin{aligned}
\sum_{\mathbf{k}} & \sum_{n \neq m} \left( e \mathbf{E} \cdot \mathbf{A}_{nm}^{(0)} \right) | \psi_{n\mathbf{k}}^{(0)}  \rangle \langle \psi_{m\mathbf{k}}^{(0)} | \\
+\sum_{\mathbf{k}} & \sum_{n,m} \left[ S_{nm} \left( \varepsilon_m^{(0)} (\mathbf{k}) + e \mathbf{E} \cdot \mathbf{A}_{mm}^{(0)}(\mathbf{k}) \right) \right. \\
& \left. -\left( \varepsilon_n^{(0)} (\mathbf{k}) + e \mathbf{E} \cdot \mathbf{A}_{nn}^{(0)}(\mathbf{k}) \right) S_{nm} \right] | \psi_{n\mathbf{k}}^{(0)} \rangle \langle \psi_{m\mathbf{k}}^{(0)} | \\
= 0 & 
\end{aligned}
\end{equation}

By checking the diagonal part, we find:
\begin{equation}
\begin{aligned}
& S_{nn} \left( \varepsilon_n^{(0)} (\mathbf{k}) + e \mathbf{E} \cdot \mathbf{A}_{nn}^{(0)}(\mathbf{k}) \right) \\
& - \left( \varepsilon_n^{(0)} (\mathbf{k}) + e \mathbf{E} \cdot \mathbf{A}_{nn}^{(0)}(\mathbf{k}) \right) S_{nn} = 0
\end{aligned}
\end{equation}
which tells us nothing. Considering the anti-Hermiticity of the \(S\) operator, we choose:
\begin{equation}
S_{nn} = 0
\end{equation}

Checking the off-diagonal part, we get:
\begin{equation}
\begin{aligned}
& e \mathbf{E} \cdot \mathbf{A}_{nm}^{(0)} + S_{nm} \left( \varepsilon_m^{(0)} (\mathbf{k}) + e \mathbf{E} \cdot \mathbf{A}_{mm}^{(0)}(\mathbf{k}) \right) \\
& - \left( \varepsilon_n^{(0)} (\mathbf{k}) + e \mathbf{E} \cdot \mathbf{A}_{nn}^{(0)}(\mathbf{k}) \right) S_{nm} = 0
\end{aligned}
\end{equation}
which yields:
\begin{equation}
\begin{aligned}
S_{nm} & = \frac{- e \mathbf{E} \cdot \mathbf{A}_{nm}^{(0)}}{\varepsilon_{nm}^{(0)} - e \mathbf{E} \cdot (\mathbf{A}_n - \mathbf{A}_m) } \\
& \approx \underbrace{ \frac{- e \mathbf{E} \cdot \mathbf{A}_{nm}^{(0)}}{\varepsilon_{nm}^{(0)}} }_{\mathcal{O}(\mathbf{E})} - \underbrace{\frac{e^2 E^a E^b A_{nm}^a (A_n^b - A_m^b)}{(\varepsilon_{nm}^{(0)})^2}}_{\mathcal{O}(\mathbf{E}^2)} + \mathcal{O}(\mathbf{E}^3) \\
& \equiv \epsilon S^{(1)}_{nm} + \epsilon^2 S_{nm}^{(2)} + \mathcal{O}(\mathbf{E}^3)
\end{aligned}
\end{equation}
where,
\begin{equation}
\varepsilon^{(0)}_{nm} \equiv \varepsilon^{(0)}_n - \varepsilon^{(0)}_m
\end{equation}

After determining the \(S\) operator, we can get the expression for \(H'\):
\begin{equation}
\begin{aligned}
H' & = H_0 + \frac{1}{2} \epsilon [S, H_1] + \frac{1}{3} \epsilon [S,[S,H_1]] + \dots \\
& = H_0 + \frac{1}{2} \epsilon^2 [S^{(1)}, H_1] \\
& + \epsilon^3 \left[ \frac{1}{2}  [S^{(2)}, H_1] + \frac{1}{3} [S^{(1)},[S^{(1)}, H_1]] \right] + \mathcal{O}(\mathbf{E}^4) \\
& =H_0 + \underbrace{H_1'}_{\mathcal{O}(\mathbf{E}^2)}
\end{aligned}
\end{equation}

Now, the unperturbed part of the SW-transformed Hamiltonian can be regarded as the diagonal part of the original Hamiltonian \(H_0\), whose unperturbed eigenstates are just \(|\psi_{n\mathbf{k}}^{(0)}\rangle\), and the corresponding eigenvalues are \((\varepsilon_n^{(0)} - e \mathbf{E} \cdot \mathbf{A}_{nn}^{(0)})\). We have transformed the perturbation Hamiltonian into a \(\mathcal{O}(\mathbf{E}^2)\) term.

Thus, the 1st order perturbation theory for this SW-transformed Hamiltonian is enough to tell us the correction of dispersion up to 3rd order, i.e., \(\varepsilon_n^{(3)}\). Since the perturbation is \(\mathcal{O}(\mathbf{E}^2)\) and contains \(\mathcal{O}(\mathbf{E}^4)\), the correction of the dispersion up to \(\varepsilon_n^{(4)}\) will contain contributions from both the 1st order perturbation theory’s \(\mathcal{O}(\mathbf{E}^4)\) term and the 2nd order perturbation theory’s \((H_1')^2 \sim \mathcal{O}(\mathbf{E}^4)\) term. However, this discussion is complex and will not be covered here. Additionally, since the SW transformation is unitary, the calculation yields the same eigenvalue for the original Hamiltonian \(H\).

\subsubsection{Correction of band dispersion}

We calculate the correction to eigenvalues order by order. The SW-transformed Hamiltonian's 0th order eigenvalue already contains \(\mathcal{O}(\mathbf{E}^0)\) and \(\mathcal{O}(\mathbf{E}^1)\) corrections for the original Hamiltonian eigenvalues:
\begin{equation}
\varepsilon_n^{(1)} = - e E^a A_{nn}^a
\end{equation}

The SW-transformed Hamiltonian's 1st order eigenvalue contains the \(\mathcal{O}(\mathbf{E}^2)\) correction, where \(\varepsilon_n^{(2)}\) is:
\begin{equation}
\begin{aligned}
\varepsilon_n^{(2)} & = \langle \psi_{n \mathbf{k}}^{(0)} | \frac{1}{2} \epsilon^2 [S^{(1)}, H_1] | \psi_{n \mathbf{k}}^{(0)} \rangle \\
& = \frac{1}{2} \sum_{m \neq n} \epsilon S_{nm}^{(1)} (- e \mathbf{E} \cdot \mathbf{A}_{mn}^{(0)} ) - (- e \mathbf{E} \cdot \mathbf{A}_{nm}^{(0)} ) \epsilon S_{mn}^{(1)} \\
& = \frac{1}{2} e^2 E^a E^b \left( \sum_{m \neq n} \frac{A_{nm}^a A_{mn}^b + A_{nm}^b A_{mn}^a}{\varepsilon_{nm}^{(0)}} \right)
\end{aligned}
\end{equation}

To simplify notation, we define the \(a\)-th component of unperturbed Berry connection as \(A_{nm}^a \equiv (\mathbf{A}_{nm}^{(0)})^a\).

\subsubsection{Correction of wave function and Berry connection}

The correction to the wave function corresponds to
\begin{equation}
\begin{aligned}
|n\rangle & \equiv e^{-S} |n'\rangle \\
& = |n'\rangle - S |n'\rangle + \frac{1}{2!} S^2 |n'\rangle + \dots \\
& = (|n^{(0)}\rangle + \mathcal{O}(\mathbf{E}^2) ) - S (|n^{(0)}\rangle + \mathcal{O}(\mathbf{E}^2) ) + \mathcal{O}(\mathbf{E}^2) \\
& = |n^{(0)}\rangle - S |n^{(0)}\rangle + \mathcal{O}(\mathbf{E}^2)
\end{aligned}
\end{equation}

Here, \(|n \rangle\) is the eigenstate of the original perturbed system (before the SW transformation), and \(|n' \rangle\) is the eigenstate of the Hamiltonian after the SW transformation. We've used the fact that \(|n'\rangle\) is \(|n^{(0)}\rangle\) up to \(\mathcal{O}(\mathbf{E}^2)\) and \(S \sim \mathcal{O}(\mathbf{E})\), where \(|n^{(0)}\rangle\) represents the eigenstate of the original unperturbed Hamiltonian.

The corrected Berry connection is:
\begin{equation}
\begin{aligned}
& \mathbf{A}_{nm} = \langle \psi_{n\mathbf{k}} | \mathbf{r} | \psi_{m \mathbf{k}} \rangle \\
& = \left( \langle \psi_{n\mathbf{k}}^{(0)} | - \langle \psi_{n\mathbf{k}}^{(0)} | S^\dagger + \mathcal{O}(\mathbf{E}^2) \right)  \mathbf{r} \left( | \psi_{n\mathbf{k}}^{(0)} \rangle - S | \psi_{n\mathbf{k}}^{(0)} \rangle + \mathcal{O}(\mathbf{E}^2) \right) \\
& = \underbrace{\mathbf{A}_{nm}^{(0)}}_{\mathcal{O}(\mathbf{E}^0)} + \underbrace{\langle \psi_{n\mathbf{k}}^{(0)} | [S, \mathbf{r}] | \psi_{m \mathbf{k}}^{(0)} \rangle}_{\mathcal{O}(\mathbf{E}^1)} + \mathcal{O}(\mathbf{E}^2)
\end{aligned}
\end{equation}

Therefore, the 1st order correction to the Berry connection is:
\begin{equation}
\begin{aligned}
& (\mathbf{A}_{nm}^{(1)})^b \\
& = \langle \psi_{n\mathbf{k}}^{(0)} | [S^{(1)}, \mathbf{r}^b] | \psi_{m \mathbf{k}}^{(0)} \rangle \\
& = \sum_{l} S_{nl}^{(1)} (\mathbf{A}_{lm}^{(0)})^b - (\mathbf{A}_{nl}^{(0)})^b S_{lm}^{(1)} \\
& = -e \mathbf{E}^a \sum_{l} \frac{ (\mathbf{A}_{nl}^{(0)})^a (\mathbf{A}_{lm}^{(0)})^b}{\varepsilon_{nl}^{(0)}} - \frac{ (\mathbf{A}_{nl}^{(0)})^b (\mathbf{A}_{lm}^{(0)})^a }{\varepsilon_{lm}^{(0)}} \\
& = -e \mathbf{E}^a G^{ab}_{nm}
\end{aligned}
\end{equation}

\(G^{ab}_{nm} \equiv \sum_{l} 
(\mathbf{A}_{nl}^{(0)})^a (\mathbf{A}_{lm}^{(0)})^b/\varepsilon_{nl}^{(0)} - (\mathbf{A}_{nl}^{(0)})^b (\mathbf{A}_{lm}^{(0)})^a/\varepsilon_{lm}^{(0)}\) is defined as the (non-Abelian) band-normalized quantum metric. For response up to 2nd order, we only need its diagonal (Abelian) part, i.e., the Band-normalized Quantum metric of the \(n\)-th band:
\begin{equation}
G_n^{ab} \equiv \sum_{m \neq n} \frac{ (\mathbf{A}_{nm}^{(0)})^a (\mathbf{A}_{mn}^{(0)})^b + (\mathbf{A}_{nm}^{(0)})^b (\mathbf{A}_{mn}^{(0)})^a }{\varepsilon_{nm}^{(0)}}
\end{equation}

The corresponding 1st order correction of Berry curvature is:
\begin{equation}
(\mathbf{\Omega}^{(1)}_n)^c = \epsilon_{abc} \partial_a (\mathbf{A}^{(1)}_{nn}) = - e \mathbf{E}^d \epsilon_{abc} \partial_a G_n^{db}
\end{equation}

\subsection{2nd order current and conductivity}\label{secconductivity}

Now, with all the ingredients, we can calculate the explicit form of the 2nd-order current and conductivity:
\begin{equation}
\mathbf{j}^{(2)}=-e \int_k \sum_n f_n^{(2)} \mathbf{v}_n^{(0)}+f_n^{(1)} \mathbf{v}_n^{(1)}+f_n^{(0)} \mathbf{v}_n^{(2)}
\end{equation}
where:
\begin{equation}
f_n^{(l)} = \left(\frac{e \tau}{\hbar} \mathbf{E} \cdot \nabla \right)^l f_0
\end{equation}
\begin{equation}
\mathbf{v}_n^{(l)} = \frac{1}{\hbar} \frac{\partial \varepsilon^{(l)}}{\partial \mathbf{k}} - \frac{e}{\hbar} \mathbf{E} \times \mathbf{\Omega}_n^{(l-1)}
\end{equation}

Substituting all the above formulas into the expression for 2nd-order current yields:
\begin{equation}
\begin{aligned}
& (\mathbf{j}^{(2)})^c = \\
& -e \int_k \sum_n (\frac{e \tau}{\hbar})^2 E_a E_b (\partial_a \partial_b f_n^{(0)}) \times \frac{1}{\hbar} \partial_c \varepsilon_n^{(0)} \\
& -e \int_k \sum_n (\frac{e \tau}{\hbar}) E_a (\partial_a f_n^{(0)}) \times ( \frac{1}{\hbar} \partial_c \varepsilon_n^{(1)} - \frac{e}{\hbar} \epsilon_{cbd} E_b (\mathbf{\Omega}_n^{(0)})^d )\\
& -e \int_k \sum_n f_n^{(0)} \times ( \frac{1}{\hbar} \partial_c \varepsilon_n^{(2)} - \frac{e}{\hbar} \epsilon_{cbd} E_b (\mathbf{\Omega}_n^{(1)})^d )\\
\end{aligned}
\end{equation}

We can change the vector form of Abelian Berry curvature into the antisymmetric matrix form, i.e.,
\((\mathbf{\Omega}_n^{(0)})^d \equiv \frac{1}{2} \epsilon_{def} (\Omega_n^{(0)})^{ef}\), and we can also use the identity \(\epsilon_{abc} \epsilon_{cde} = \delta_{ad} \delta_{be} - \delta_{ae} \delta_{bd}\) and integration by parts, further symmetrizing the expression with respect to \(a\), \(b\) indices, which gives:
\begin{equation}
\begin{aligned}
& (\mathbf{j}^{(2)})^c = \\
& - \frac{e^2 \tau^2}{\hbar^3} E_a E_b \sum_n \int_k f_n^{(0)} \partial_a \partial_b \partial_c \varepsilon_n^{(0)} \\
& + \frac{e^3 \tau}{\hbar^2} E_a E_b \sum_n \int_k f_n^{(0)} \frac{1}{2} (\partial_a \Omega_n^{bc} + \partial_b \Omega_n^{ac}) \\
& - \frac{e^3}{\hbar} E_a E_b \sum_n \int_k f_n^{(0)} \left[ 2 \partial_c G_n^{ab} - \frac{1}{2}( \partial_a G_n^{bc} + \partial_b G_n^{ac} ) \right]
\end{aligned}
\end{equation}

Comparing with the definition of 2nd-order conductivity \((\mathbf{j}^{(2)})^c = \sigma^{ab;c} E_a E_b\), we can get the expression for 2nd-order conductivity:
\begin{equation}
\begin{aligned}
& \sigma^{ab;c} \equiv \sigma^{ab;c}_{disp} + \sigma^{ab;c}_{BCD} + \sigma^{ab;c}_{geo} = \\
& - \frac{e^2 \tau^2}{\hbar^3} \sum_n \int_k f_n^{(0)} \partial_a \partial_b \partial_c \varepsilon_n^{(0)} \\
& + \frac{e^3 \tau}{\hbar^2} \sum_n \int_k f_n^{(0)} \frac{1}{2} (\partial_a \Omega_n^{bc} + \partial_b \Omega_n^{ac}) \\
& - \frac{e^3}{\hbar} \sum_n \int_k f_n^{(0)} \left[ 2 \partial_c G_n^{ab} - \frac{1}{2}( \partial_a G_n^{bc} + \partial_b G_n^{ac} ) \right]
\end{aligned}\label{eqS}
\end{equation}

Here, the second line of \(\sigma^{ab;c}\) is a Drude-like term and corresponds to a longitudinal response. It's completely determined by the dispersion of Bloch states and proportional to \(\tau^2\), making it vanish under \(\mathcal{T}\)-symmetry.

The third line is the famous Berry Curvature Dipole (BCD) term proposed in \cite{sodemann2015quantum} and corresponds to a transverse response. It's a topological response and proportional to \(\tau^1\), making it nonvanishing even when \(\mathcal{T}\)-symmetry is present. Specifically, this term becomes nonzero in \(\mathcal{P}\)-symmetry-breaking systems and leads to the nonlinear anomalous Hall effect (NLAHE).

The fourth term is the (Band-normalized) quantum metric dipole term and is a mixed response with both transverse and longitudinal contributions. It's a geometric response and proportional to \(\tau^0\), making it independent of the phenomenological relaxation time \(\tau\), and thus is termed "intrinsic".

%

\end{document}